\documentclass[12pt]{iopart}

\usepackage{iopams} 
\usepackage{bm}
\usepackage{graphicx}
\usepackage{hyperref}

\eqnobysec

\begin{document}

\title[Close-limit approximation with post-Newtonian initial conditions]{The Close-Limit Approximation for Black Hole Binaries with Post-Newtonian Initial Conditions}

\author{Alexandre Le Tiec and Luc Blanchet}

\ead{letiec@iap.fr, blanchet@iap.fr}

\address{$\mathcal{G}\mathbb{R}\varepsilon{\mathbb{C}}\mathcal{O}$, Institut d'Astrophysique de Paris --- UMR 7095 du CNRS, Universit\'e Pierre et Marie Curie, 98$^{\rm bis}$ boulevard Arago, 75014 Paris, France}

\begin{abstract}
The ringdown phase of a black hole formed from the merger of two orbiting black holes is described by means of the close-limit (CL) approximation starting from second-post-Newtonian (2PN) initial conditions. The 2PN metric of point-particle binaries is formally expanded in CL form and identified with that of a perturbed Schwarzschild black hole. The multipolar coefficients describing the even-parity (or polar) and odd-parity (axial) components of the linear perturbation consistently satisfy the 2PN-accurate perturbative field equations. We use these coefficients to build initial conditions for the Regge-Wheeler and Zerilli wave equations, which we then evolve numerically. The ringdown waveform is obtained in two cases: head-on collision with zero-angular momentum, composed only of even modes, and circular orbits, for which both even and odd modes contribute. In a separate work, this formalism is applied to the study of the gravitational recoil produced during the ringdown phase of coalescing binary black holes.
\end{abstract}

\pacs{04.25.Nx, 04.30.-w, 97.60.Lf}

\maketitle

\section{Introduction}\label{secI}

Post-Newtonian (PN) methods have proved to be extremely efficient in describing the inspiral phase of compact binary systems, up to about the location of the innermost circular orbit (ICO). The PN inspiral signal has been developed up to 3.5PN order\footnote{As usual the $n$PN order refers either to the terms $\sim 1/c^{2n}$ in the equations of motion, with respect to the usual Newtonian acceleration, or in the radiation field, relatively to the standard quadrupolar waveform.} for the orbital phase evolution \cite{Bl.al.02,Bl2.al.04} and up to 3PN order \cite{Bl.al.96,Ar.al.04,Bl.al.08} in the amplitude waveform (see \cite{Bl.06} for a review). On the other hand, recent advances in numerical calculations of binary black holes \cite{Pr.05,Ca.al.06,Ba.al.06} have provided a very accurate description of the subsequent merger and ringdown phases, say, from the ICO on. The comparison of the numerical-relativity and PN results is a crucial task that has been successfully achieved \cite{Bu.al.07,Ba.al.07,Be.al.07,Pa.al.07,Ha.al.07}. Their matching is currently under way \cite{Aj.al.08} and should yield a complete and very accurate solution of the problem of binary coalescence.

Nevertheless, analytic and/or semi-analytic methods are still very useful for gaining more physical understanding of the relaxation of binary black holes toward their final equilibrium state (see e.g. \cite{Ke.al.09} for a recent example). Of particular importance is the close-limit (CL) approximation method, pioneered by Price \& Pullin \cite{PrPu.94} and Abrahams \& Price \cite{AbPr.96}. The CL approximation permits the description of the last stage of evolution of a black hole binary, when the two black holes are close enough that they are surrounded by a common horizon, as a perturbation of a single (Schwarzschild or Kerr) black hole. Recent revisits of the CL approach made use of numerically generated initial data \cite{Ba.al.02}, and Bowen-York-type initial conditions \cite{So.al.06,So.al.07}. Alternative analytic or semi-analytic schemes for dealing with the same problem are based on the effective-one-body approach \cite{BuDa.99} (see \cite{DaNa.09} for a recent review).

In the present paper we shall implement the CL approximation starting from post-Newtonian initial conditions, appropriate for the initial inspiral phase of binary black holes. A physical motivation is that the results of numerical relativity \cite{Pr.05,Ca.al.06,Ba.al.06} show that the pulse of radiation coming from the merger phase is very short and seems to connect smoothly to the previous inspiral and subsequent ringdown phases. It is thus reasonable to expect that PN initial conditions starting the CL evolution should essentially yield the right physics for the ringdown phase. The application of this formalism to the computation of the gravitational recoil effect or ``kick'' occuring during the ringdown phase will be presented in a separate work \cite{Le.al.09}.

Let us outline the method. We conveniently distinguish several dimensionless ratios to describe a compact binary system. First, we introduce the ``post-Minkowskian'' (PM) ratio, measuring the internal gravity responsible for the dynamics of the system, and defined by
\begin{equation}\label{epsPM}
	\varepsilon_{\rm PM} \sim \frac{G M}{c^2 r_{12}} \, ,
\end{equation}	
where $r_{12}$ is the typical distance between the two compact bodies, and $M=m_1+m_2$ is the sum of their masses. Second, a post-Newtonian expansion will essentially be an expansion in powers of the \textit{a priori} distinct slowness parameter
\begin{equation}\label{epsPN}
	\varepsilon_{\rm PN} \sim \frac{v_{12}^2}{c^2} \, ,
\end{equation}	
where $v_{12}$ is the typical value of the orbital relative velocity. Recall that the PN expansion is only valid in the near zone defined by $r \ll \lambda$, where $\lambda \sim r_{12} / \sqrt{\varepsilon_{\rm PM}}$ is the typical wavelength of the emitted gravitational waves, and $r$ the distance from the field point to, say, the center of mass of the binary.

For a binary system moving on a circular orbit the two parameters $\varepsilon_{\rm PN}$ and $\varepsilon_{\rm PM}$ are comparable, $\varepsilon_{\rm PN}\sim \varepsilon_{\rm PM}$. In this case, if we limit the PN expansion to a few terms, we need $\varepsilon_{\rm PN} \ll 1$ hence $r_{12} \gg G M/c^2$. However, it is often better to view $\varepsilon_{\rm PN}$ and $\varepsilon_{\rm PM}$ as independent parameters because if the binary system is moving on a highly eccentric bound orbit with eccentricity $e \lesssim 1$, the PN parameter can be much smaller that the PM one, since at the apoapsis of the orbit we have $\varepsilon_{\rm PN}\sim (1-e) \, \varepsilon_{\rm PM} \ll \varepsilon_{\rm PM}$. And, for an unbound orbit with eccentricity $e\gg 1$, we would have $\varepsilon_{\rm PN}\sim (1+e) \, \varepsilon_{\rm PM} \gg \varepsilon_{\rm PM}$ at the periapsis.

On the other hand, the close-limit approximation consists of an expansion in powers of the dimensionless ratio considered small\footnote{In the works \cite{PrPu.94,AbPr.96}, the CL parameter is defined as $c^2 r_{12}/ (G M)$. More recently, Sopuerta {\em et al.} \cite{So.al.06} adopted the definition \eref{epsCL}. In the formal limit $r_{12} \rightarrow 0$, these two definitions are equivalent.}
\begin{equation}\label{epsCL}
	\varepsilon_{\rm CL} \sim \frac{r_{12}}{r} \, .
\end{equation}	
This expansion can formally be viewed as an expansion when the size of the source tends to zero, or multipolar expansion. Therefore, if we limit the expansion to a few terms, we need $\varepsilon_{\rm CL} \ll 1$ and the CL approximation is expected to be valid in the domain $r \gg r_{12}$ (like a multipole expansion).

Clearly the PN and CL approximations that we intend to employ simultaneously have disconnected domains of validity. Indeed, the CL describes a slightly distorted black hole such that $r_{12} \gtrsim G M/c^2$, so that for circular orbits $\varepsilon_{\rm PN} \lesssim 1$, which makes the near zone very small; in other words, the PN metric will only be valid very close to the source while the CL approximation requires $r \gg r_{12}$. Despite such apparent clash, we shall be inspired by the method of matched asymptotic expansions \cite{Lag}, which in principle allows one to get an analytic expression valid in the entire domain $0 \leqslant r < +\infty$. Of course, this method is based on the existence of an overlapping zone, where the two asymptotic expansions are simultaneously valid and can be matched together. But in the present context there is no such thing as a overlapping zone. Hence our use of the theory of matched asymptotic expansions to relate PN and CL approximations can at best be only \textit{formal}.

Starting from the PN metric, already in the form of an expansion in powers of $\varepsilon_{\rm PN}$, we shall restrict ourselves to the terms linear in $\varepsilon_{\rm PM}$ (i.e., essentially, linear in $G$). This is to be consistent later with the use of a linear black hole perturbation. Then, each of the coefficients of the PN metric will be expanded in powers of $\varepsilon_{\rm CL}$, which will enable us to identify the Schwarzschild background metric (up to terms of order $\varepsilon_{\rm PM}^2$) and the perturbation $h_{\mu \nu}$ of that background. Thus the perturbation will appear as a double expansion series of the type [cf. the explicit results \eref{metricPNCL00}--\eref{metricPNCLij}]
\begin{equation}\label{double_exp}
	h_{\mu \nu} = \varepsilon_{\rm PM} \sum_{n\geqslant0} \sum_{k\geqslant0} h_{\mu \nu}^{(n,k)} \, \varepsilon_{\rm PN}^n \, \varepsilon_{\rm CL}^{k+1} + \mathcal{O}(\varepsilon_{\rm PM}^2) \, ,
\end{equation}
where $n$ refers to the post-Newtonian order and $k$ can be viewed as the multipolar order of the expansion.\footnote{Our convention is that $k$ represents the power of the separation $r_{12}$ in the CL expansion, taking into account the inverse power of $r_{12}$ hidden in the PM indicator $\varepsilon_{\rm PM}$ in front of \eref{double_exp}.} In principle, one could perform the expansions in the opposite way, i.e. expanding first in powers of $\varepsilon_{\rm CL}$, and then in powers of $\varepsilon_{\rm PN}$. In the method of matched asymptotic expansions the result should be the same, i.e. term by term identical in the double expansion series. This would however require first the knowledge of the black hole perturbation metric in the CL approximation; such metric can only be computed numerically. 

In the present paper, we shall implement the expansion \eref{double_exp}, limiting ourselves to second post-Newtonian order. The reason is that the metric is needed in closed analytic form for any field point in the near zone, and that the 3PN metric is currently not known for any field point; only the 3PN metric when regularized at the very location of the particles is known \cite{Bl.al.09}. One of our aims is the study reported in the separate work \cite{Le.al.09} of the gravitational recoil effect. The recoil is the reaction of the binary system to the linear momentum carried away by the gravitational waves, and results at leading order from the interaction between the $\ell=2$ and the $\ell=3$ modes, where $\ell$ is the azimuthal number of the decomposition of the black hole perturbation onto tensorial spherical harmonics; we shall thus push the CL expansion up to at least octupolar order, i.e. $k\geqslant3$, to ensure that the modes $\ell=2$ and $3$ are both taken into account.

The present approach will be limited to the case of a slowly spinning black hole. The initial orbital angular momentum of the binary system, which is constant and supposed to give the spin of the final black hole, is considered as part of the perturbation of a Schwarzschild black hole, and is necessarily small. However we have learned from numerical calculations that the final black hole produced by coalescence is a rapidly spinning Kerr black hole \cite{Pr.05,Ca.al.06,Ba.al.06}. The ringdown waveform that we shall compute in this paper will be that of a perturbed Schwarzschild black hole, and hence the quasi-normal mode frequencies will not include the effect of the black hole spin. To remedy this problem and get better agreement with numerical relativity would necessitate similar calculations using a Kerr black hole background.

The remainder of this paper is organized as follows: In Sec.~\ref{secII} we consider the 2PN metric of two compact bodies at first post-Minkowskian order (to be consistent with first-order black hole perturbations) and formally re-expand it in the CL form. In Sec.~\ref{secIII} we first give a short recap of the theory of linear perturbations of a Schwarzschild black hole, and then use this formalism to identify the perturbation associated with the previously CL-expanded 2PN metric. In Sec.~\ref{secIV} we verify that the field equations for this metric are satisfied. Using the CL-expanded 2PN metric as initial data, we numerically evolve the Regge-Wheeler and Zerilli functions in Sec.~\ref{secV}, and present the resulting waveforms generated during the ringdown phase of coalescing black holes, for both even and odd-parity perturbations. Finally we conclude in Sec.~\ref{secVI}. Some necessary details on black hole perturbation theory are relagated to \ref{appA}.

\section{The 2PN metric in close-limit form}\label{secII}

\subsection{The 2PN metric for two point masses}\label{secIIsubI}

In the present paper we shall solve numerically the Regge-Wheeler and Zerilli wave equations [see Eqs.~\eref{ZRW} below] starting from post-Newtonian initial conditions. Thus we assume that the initial metric at the end of the inspiral phase is given by the standard PN metric generated by two point masses $m_1$ and $m_2$ modelling two non-spinning black holes. We adopt the 2PN precision because the 3PN metric in the near zone is not known in the ``bulk'', i.e. for any field point outside the position of the particles.

\begin{figure}
	\begin{center}
		\includegraphics[width=12cm]{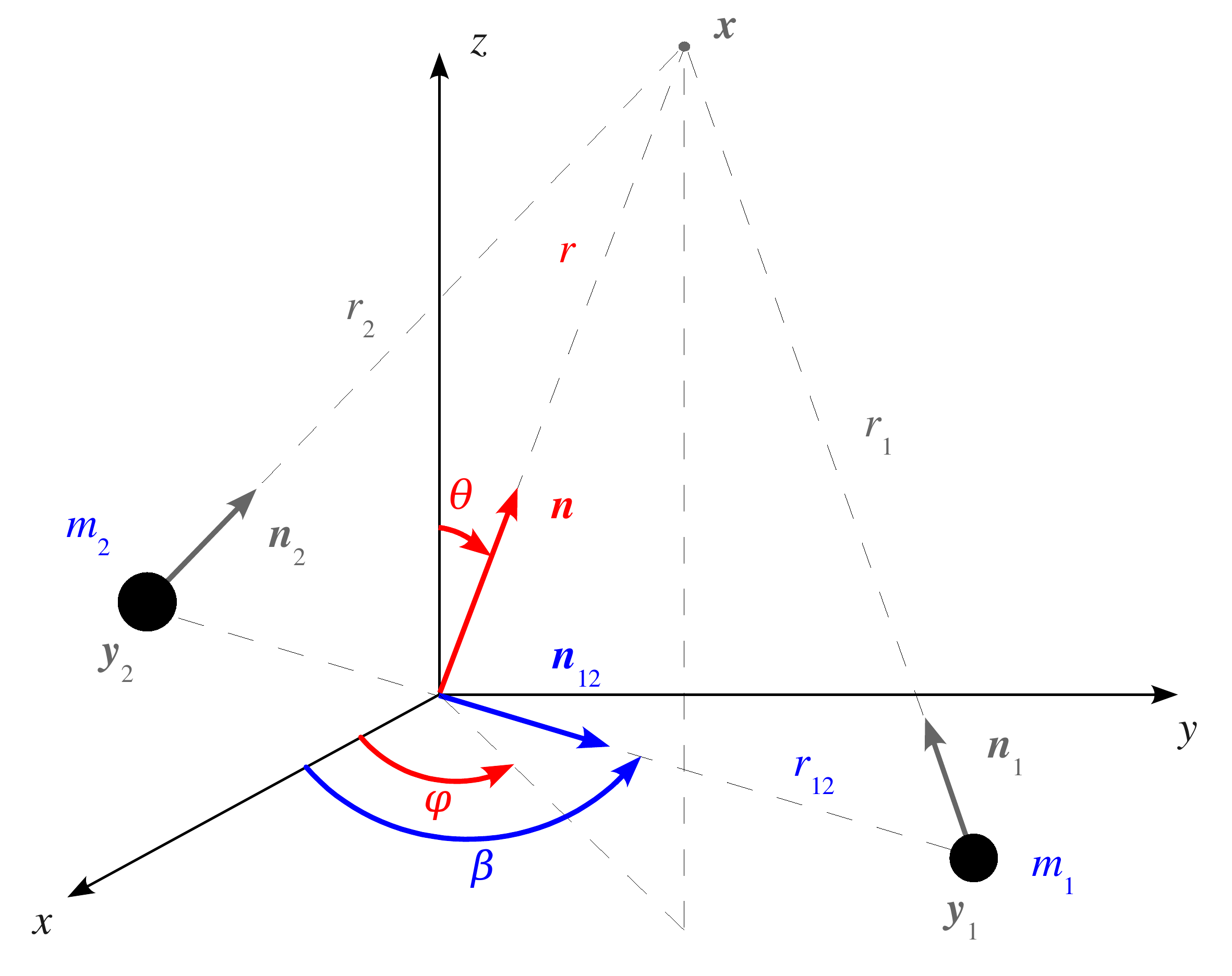}
		\caption{\footnotesize Two black holes of Schwarzschild masses $m_{1,2}$ in a Cartesian coordinate system $x^\mu=\{c t,x,y,z\}$ centered on the binary's center of mass, or the associated spherical coordinate system $x^\alpha=\{c t,r,\theta,\varphi\}$.}
		\label{fig_binary}
	\end{center}
\end{figure}

Our calculation starts with the post-Newtonian metric $G_{\mu\nu}^{\rm PN}$ written in a Cartesian harmonic coordinate system, and given as \cite{Bl.al.98}\footnote{Greek indices take space-time values $0,1,2,3$. The indices $\mu, \nu, \dots$ indicate Cartesian coordinates $x^\mu=\{c t,x,y,z\}$, while $\alpha, \beta, \dots$ refer to spherical coordinates $x^\alpha=\{c t,r,\theta,\varphi\}$. Latin indices $i, j, \dots$ take spatial values $1,2,3$. Bold-face notation is often used to denote ordinary spatial vectors, $\mathbf{x}=\{x^i\}$. The two black holes are often labeled by $A=1,2$. Parentheses around indices are used to indicate symmetrisation, e.g. $U^{(i} V^{j)} = \frac{1}{2} \left( U^i V^j + U^j V^i\right)$. The usual (Euclidean) scalar product between two 3-vectors $\mathbf{U}$ and $\mathbf{V}$ is denoted $(U V)$, e.g. $(n_1 v_1) = \mathbf{n}_1 \cdot \mathbf{v}_1$. To the terms given below in Eqs.~\eref{metricGharm00}--\eref{metricGharmij}, we have to add those ones corresponding to the relabeling $1 \leftrightarrow 2$ (with the obvious exception of the Minkowski metric which should not be counted twice).}
\numparts
	\begin{eqnarray}
	\fl G_{00}^{\rm PN} = -1 + \frac{2 G m_1}{c^2 r_1} + \frac{1}{c^4} \Biggl[ \frac{G m_1}{r_1} \left( - (n_1v_1)^2 + 4 v_1^2 \right) - 2 \frac{G^2 m_1^2}{r_1^2} \nonumber \\+ \, G^2 m_1 m_2 \left( - \frac{2}{r_1 r_2} - \frac{r_1}{2 r_{12}^3} + \frac{r_1^2}{2 r_2 r_{12}^3} - \frac{5}{2 r_2 r_{12}} \right) \Biggr] \nonumber \\+ \, \frac{4 G^2 m_1 m_2}{3 c^5 r_{12}^2} (n_{12}v_{12}) + 1 \leftrightarrow 2 + \mathcal{O}(c^{-6}) \, , \label{metricGharm00} \\
%%%%%%%%%%%%%%%%%%%%%%%%%%%%%%%%%%%%%%%%%%%%%%%%%%%%%%%%
	\fl G_{0i}^{\rm PN} = - 4 \frac{G m_1}{c^3 r_1} v_1^i + \frac{1}{c^5} \Biggl[ n_1^i \Biggl( - \frac{G^2 m_1^2}{r_1^2} (n_1v_1) + \frac{G^2 m_1 m_2}{S^2} \biggl( - 16 (n_{12}v_1) + 12 (n_{12}v_2) \nonumber \\ \qquad \qquad \quad - 16 (n_2v_1) + 12 (n_2v_2) \biggr) \Biggr) \nonumber \\ + \, n_{12}^i G^2 m_1 m_2 \Biggl( - 6 (n_{12}v_{12}) \frac{r_1}{r_{12}^3} - 4 (n_1v_1)\frac{1}{r_{12}^2} + 12 (n_1v_1)\frac{1}{S^2} \nonumber \\ \qquad \qquad \qquad - 16 (n_1v_2) \frac{1}{S^2} + 4 (n_{12}v_1) \frac{1}{S} \Biggl( \frac{1}{S} + \frac{1}{r_{12}} \Biggr) \Biggr) \nonumber \\ + \, v_1^i \Biggl( \frac{G m_1}{r_1} \left( 2 (n_1v_1)^2 - 4 v_1^2 \right) + \frac{G^2 m_1^2}{r_1^2} + G^2 m_1 m_2 \left( \frac{3 r_1}{r_{12}^3} - \frac{2 r_2}{r_{12}^3} \right) \nonumber \\ \qquad + G^2 m_1 m_2  \left( - \frac{r_2^2}{r_1 r_{12}^3} - \frac{3}{r_1 r_{12}} + \frac{8}{r_2 r_{12}} - \frac{4}{r_{12} S} \right) \Biggr) \Biggr] \nonumber \\ + 1 \leftrightarrow 2 + \mathcal{O}(c^{-6}) \ , \\
%%%%%%%%%%%%%%%%%%%%%%%%%%%%%%%%%%%%%%%%%%%%%%%%%%%%%%%%%%%
	\fl G_{ij}^{\rm PN} = \delta^{ij} + \frac{2 G m_1}{c^2 r_1} \delta^{ij} + \frac{1}{c^4} \Biggl[ \delta^{ij} \Biggl( - \frac{G m_1}{r_1} (n_1v_1)^2 + \frac{G^2 m_1^2}{r_1^2} \nonumber \\ \qquad \qquad \qquad + G^2 m_1 m_2 \left( \frac{2}{r_1 r_2} - \frac{r_1}{2 r_{12}^3} + \frac{r_1^2}{2 r_2 r_{12}^3} - \frac{5}{2 r_1 r_{12}} + \frac{4}{r_{12} S} \right) \Biggr) \nonumber \\ + 4 \frac{G m_1}{r_1} v_1^i v_1^j+\frac{G^2 m_1^2}{r_1^2} n_1^i n_1^j - 4 G^2 m_1 m_2 n_{12}^i n_{12}^j \left( \frac{1}{S^2} + \frac{1}{r_{12} S} \right) \nonumber \\ \qquad \qquad + \frac{4 G^2 m_1 m_2}{S^2} \left( n_1^{(i} n_2^{j)} + 2 n_1^{(i} n_{12}^{j)} \right) \Biggr] \nonumber \\ + \frac{G^2 m_1 m_2}{c^5 r_{12}^2} \left( -\frac{2}{3} (n_{12}v_{12}) \delta^{ij} - 6 (n_{12}v_{12}) n_{12}^i n_{12}^j+ 8 n_{12}^{(i} v_{12}^{j)} \right) \nonumber \\ + 1 \leftrightarrow 2 + \mathcal{O}(c^{-6}) \, . \label{metricGharmij}
	\end{eqnarray}
\endnumparts
The trajectory of the $A$\textsuperscript{th} black hole is denoted $\mathbf{y}_A$ and its ordinary velocity is $\mathbf{v}_A = \rmd \mathbf{y}_A/\rmd t$, where $t=x^0/c$ is the harmonic-coordinate time. The relative velocity is denoted $\mathbf{v}_{12} = \mathbf{v}_1 - \mathbf{v}_2$. The Euclidean distance between the black hole $A$ and any field point is $r_A = \vert \mathbf{x} - \mathbf{y}_A \vert$. The unit vector pointing from $A$ to the field point is $\mathbf{n}_A = (\mathbf{x} - \mathbf{y}_A) / r_A$, and the unit direction from body 2 to body 1 reads $\mathbf{n}_{12}=\mathbf{y}_{12}/r_{12}$, where $\mathbf{y}_{12}=\mathbf{y}_{1}-\mathbf{y}_{2}$ and the binary's separation is denoted $r_{12} = \vert \mathbf{y}_1 - \mathbf{y}_2 \vert$. Several terms in \eref{metricGharm00}--\eref{metricGharmij} involve the particular combination $S\equiv r_1+r_2+r_{12}$. All these conventions can be visualized in Fig.~\ref{fig_binary}.

In the following we shall restrict ourselves to those terms in the full 2PN metric \eref{metricGharm00}--\eref{metricGharmij} which are \textit{linear} in the parameter $\varepsilon_{\rm PM}$ given by \eref{epsPM}, or equivalently in the gravitational constant $G$.\footnote{From now on it will be simpler to forget about the dimensionless estimates $\varepsilon_{\rm PM}$, $\varepsilon_{\rm PN}$ and $\varepsilon_{\rm CL}$ defined in the Introduction for pedagogical reasons. We shall replace them by the dimensionful but more obvious constants $G$ and $c^{-2}$, and parameter $r_{12}$, respectively.} Indeed, our work will be based on the theory of first-order perturbations of a Schwarzschild black hole, and the corresponding terms in the PN framework will necessarily have to involve only linear powers of $G$ to be consistent. Higher powers of $G$ in the PN metric will correspond to higher-order perturbation theory. Since the part of the 2PN metric which is linear in $G$ is obviously a solution of the Einstein field equations at 2PN order [up to terms $\mathcal{O}(G^2)$], the multipolar coefficients describing the perturbation in the CL approximation will satisfy the perturbative Einstein field equations, as checked in Sec.~\ref{secIV}. This restriction to the terms linear in $G$ in the PN metric appears therefore as necessary; the price we pay is that our initial conditions will not contain the full information encoded into the 2PN metric: for instance, all the terms involving $S$ in \eref{metricGharm00}--\eref{metricGharmij} disappear. To include meaningfully the complete 2PN metric would require using the theory of second-order perturbations of a Schwarzschild black hole \cite{Bri.al.09}. 

Neglecting the non-linear terms in $G$ we end up with a comparatively much simpler metric, reading\footnote{The remainder $\mathcal{O}(G^2,c^{-6})$ includes all terms which are at least of order $G^2$ \textit{or} of order $c^{-6}$ \textit{or} both. Thus for instance the ``radiation-reaction'' terms present at order $c^{-5}$ in $G^{\rm PN}_{00}$ and $G^{\rm PN}_{ij}$ are included in this remainder because they are also of order $G^2$.}
\numparts
	\begin{eqnarray}
		\fl G_{00}^{\rm PN} = -1 + \frac{2 G m_1}{c^2 r_1} \left[ 1 + \frac{1}{c^2} \left( 2 v_1^2 - \frac{1}{2} (n_1 v_1)^2 \right) \right] + 1 \leftrightarrow 2 + \mathcal{O}(G^2,c^{-6}) \, , \label{metricG00} \\
		\fl G_{0i}^{\rm PN} = - \frac{4 G m_1}{c^3 r_1} \left[ 1 + \frac{1}{c^2} \left( v_1^2 - \frac{1}{2} (n_1 v_1)^2 \right) \right] v_1^i + 1 \leftrightarrow 2 + \mathcal{O}(G^2,c^{-6}) \, , \\
		\fl G_{ij}^{\rm PN} = \left[ 1 + \frac{2 G m_1}{c^2 r_1} - \frac{G m_1}{c^4 r_1} (n_1 v_1)^2 \right] \delta^{ij} + \frac{4 G m_1}{c^4 r_1} v_1^i v_1^j + 1 \leftrightarrow 2 + \mathcal{O}(G^2,c^{-6}) \label{metricGij} \, .
	\end{eqnarray}
\endnumparts
At this stage we could proceed with the CL expansion to identify the Schwarzschild background metric and the perturbation. However the PN metric is in harmonic coordinates so we would obtain the Schwarzschild metric in harmonic coordinates; this is not convenient because the perturbation formalism is usually defined in standard Schwarzschild-Droste coordinates. We shall thus perform a suitable coordinate transformation such that after expanding the metric in the CL approximation we obtain directly the Schwarzschild background metric in Schwarzschild coordinates.

Since we are working at linear order in $G$ it is sufficient to perform a linear gauge transformation at order $G$, say $\delta x^\mu=\xi^\mu$, where the gauge vector is $\xi^\mu=\mathcal{O}(G)$. Note that this gauge transformation is defined with respect to the Minkowski background. Later, when using black-hole perturbations, we shall perform a gauge transformation with respect to the Schwarzschild background. A suitable gauge transformation which fulfills our purpose is then
\numparts
	\begin{eqnarray}
		\xi^{0} & = - \frac{G m_1}{c^3} \,(n_1 v_1) - \frac{G m_2}{c^3} \,(n_2 v_2) \, , \label{xi0} \\
		\xi^{i} & = \frac{G m_1}{c^2} \,n_1^i + \frac{G m_2}{c^2} \,n_2^i \, . \label{xii}
	\end{eqnarray}
\endnumparts
Such gauge transformation does \textit{not} satisfy the harmonic gauge condition $\Box \xi^\mu = 0$, therefore our new coordinates will not be harmonic. Under this gauge transformation we have
\begin{equation}\label{Gg}
		G^{\rm PN}_{\mu\nu}(x)=g^{\rm PN}_{\mu\nu}(x)+2\partial_{(\mu}\xi_{\nu)}+\mathcal{O}(G^2)\, ,
\end{equation}
where $\xi_\mu\equiv\eta_{\mu\nu}\xi^\nu$, and the new metric components read now
\numparts
	\begin{eqnarray}
		\fl g_{00}^{\rm PN} = -1 + \frac{2 G m_1}{c^2 r_1} \left[ 1 + \frac{1}{c^2} \left( 3 v_1^2 - \frac{3}{2} (n_1 v_1)^2 \right) \right] + 1 \leftrightarrow 2 + \mathcal{O}(G^2,c^{-6}) \, , \label{metric2PN00} \\
		\fl g_{0i}^{\rm PN} = - \frac{4 G m_1}{c^3 r_1} \left[ 1 + \frac{1}{c^2} \left( v_1^2 - \frac{1}{2} (n_1 v_1)^2 \right) \right] v_1^i + 1 \leftrightarrow 2 + \mathcal{O}(G^2,c^{-6}) \, , \\
		\fl g_{ij}^{\rm PN} = \delta^{ij} + \frac{2 G m_1}{c^2 r_1} \left[ n_1^i n_1^j + \frac{1}{c^2} \left( 2 v_1^i v_1^j - \frac{1}{2} (n_1 v_1)^2 \delta^{ij} \right) \right]+ 1 \leftrightarrow 2 + \mathcal{O}(G^2,c^{-6}) \, . \label{metric2PNij}
	\end{eqnarray}
\endnumparts
We shall start our perturbative CL setup from that PN metric.

It is important to comment on the post-Newtonian counting we are adopting for the 2PN metric in \eref{metric2PN00}--\eref{metric2PNij}. The accuracy of this metric is really 2PN only for the geodesic motion of ``photons'' rather than of massive particles. Indeed, we did not include here the term of order $\mathcal{O}(c^{-6})$ in the 00 component of the metric, although it is known from \cite{Bl.al.98}. This term would be needed for describing the 2PN motion of massive bodies. It will turn out to be essential to expand both $g^{\rm PN}_{00}$ and $g^{\rm PN}_{ij}$ at the same post-Newtonian order ---namely up to $\sim c^{-4}$ for both components in \eref{metric2PN00} and \eref{metric2PNij}--- because only then can we be consistent with the linear black hole perturbation. Physically this results from the fact that the CL approximation is assuming that the metric is a small deformation of that of a black hole, therefore when expanding \eref{metric2PN00}--\eref{metric2PNij} in the CL form we shall have $r_{12} \gtrsim G M/c^2$, so the orbital velocities are going to be very relativistic, i.e. $\vert \mathbf{v}_A \vert \lesssim c$, and thus the $g^{\rm PN}_{00}$, $g^{\rm PN}_{ij}$ and $g^{\rm PN}_{0i}$ components should give similar contributions to the line element $\rmd s^2$, and are therefore to be expanded up to the same PN order. This is thus similar to the accuracy needed for the geodesic motion of a photon where all the metric coefficients $g^{\rm PN}_{00}$, $g^{\rm PN}_{0i}$ and $g^{\rm PN}_{ij}$ are to be given with the same PN accuracy. On the other hand, we shall see later that it is very important to include the gravitomagnetic field $g^{\rm PN}_{0i}$ up to order $c^{-5}$, because it will play a crucial role in generating the odd-parity or \textit{axial} contributions to the perturbations.

\subsection{Close-limit expansion of the 2PN metric}\label{secIIsubII}

In the CL approximation, we expand the 2PN metric \eref{metric2PN00}--\eref{metric2PNij} in powers of the parameter \eref{epsCL}, or equivalently of the relative distance $r_{12} = \vert \mathbf{y}_1 - \mathbf{y}_2 \vert$ when $r_{12}\rightarrow 0$. To do so we shall first express all variables in the frame of the center of mass defined at the required 1PN accuracy. The individual positions $\mathbf{y}_A$ of the particles in the center-of-mass frame are given in terms of their relative position $\mathbf{y}_{12} = \mathbf{y}_1 - \mathbf{y}_2$ and read at 1PN order \cite{BlIy.03}
\begin{eqnarray}\label{rel_pos}
	\mathbf{y}_1 = \left[ X_2 + \frac{\nu}{2 c^2} \,\delta X \left( v_{12}^2 - \frac{G M}{r_{12}} \right) \right] \mathbf{y}_{12} + \mathcal{O}(c^{-4}) \, ,
\end{eqnarray}
together with $1 \leftrightarrow 2$ for the other particle. We introduced the total mass $M \equiv m_1 + m_2$, the relative mass ratios $X_A\equiv m_A/M$, and the symmetric mass ratio $\nu \equiv X_1 X_2 = m_1 m_2/M^2$, such that $0 < \nu \leqslant \frac{1}{4}$, with $\nu = \frac{1}{4}$ for an equal-mass binary, and $\nu \rightarrow 0$ in the test-particle limit. We denote also the mass difference by $\delta M \equiv m_1 - m_2$, and the relative mass difference by $\delta X \equiv X_1 - X_2 = \delta M/M$, which can also be written in terms of the symmetric mass ratio as $\delta X = \pm \sqrt{1 - 4 \nu}$. As previously the relative velocity is $\mathbf{v}_{12} = \rmd \mathbf{y}_{12} / \rmd t = \mathbf{v}_1 - \mathbf{v}_2$, and $v_{12}^2 = (v_{12} v_{12})$. Note that the 1PN correction in \eref{rel_pos} vanishes for circular orbits. Similarly, from the time derivatives of \eref{rel_pos} we get the 1PN-accurate expression of the individual velocities as
\begin{eqnarray}\label{rel_vel}
	\fl \mathbf{v}_1 = \left[ X_2 + \frac{\nu}{2 c^2} \,\delta X \left( v_{12}^2 - \frac{G M}{r_{12}} \right) \right] \mathbf{v}_{12} - \frac{\nu}{2 c^2} \frac{G \delta M}{r_{12}^2} (n_{12} v_{12}) \, \mathbf{y}_{12} + \mathcal{O}(c^{-4}) \, .
\end{eqnarray}

The expressions \eref{rel_pos}--\eref{rel_vel} were derived in harmonic gauge, and we have still to check that they are also valid in the new gauge specified by \eref{xi0}--\eref{xii}. The center-of-mass frame is defined by the vanishing of the center of mass position $\mathbf{G}$. A generic gauge transformation $\xi^\mu$ will displace the position of the center of mass by the amount $\delta_\xi \mathbf{G}=-m_1\bm{\xi}_1-m_2\bm{\xi}_2$, where $\bm{\xi}_A$ is the spatial gauge vector evaluated at the location of the particle $A$. The associated shift of the particle's individual positions with respect to the center of mass will then be given by $\delta_\xi \mathbf{y}_A=-\delta_\xi \mathbf{G}/M$ (the same for both particles). In the case of the gauge vector \eref{xi0}--\eref{xii} we readily find $\bm{\xi}_1=G m_2\,\mathbf{n}_{12}/c^2$ and $\bm{\xi}_2=-G m_1\,\mathbf{n}_{12}/c^2$, so that $\delta_\xi \mathbf{G}=\mathbf{0}$ and thus $\delta_\xi \mathbf{y}_A=\mathbf{0}$. (See more details in the Appendix B of Ref.~\cite{Bl.al.09}.)

We are now ready to write down the expansion of $r_A = \vert \mathbf{x} - \mathbf{y}_A \vert$ when the CL ratio $r_{12}/r$ tends to zero, where $r_{12} = \vert \mathbf{y}_1 - \mathbf{y}_2 \vert$ is the binary separation and $r=\vert\mathbf{x}\vert$ is the distance of the field point to the center of mass properly defined at the 1PN order. Introducing the Legendre polynomials $P_k$ and using \eref{rel_pos}, we thus have
\begin{eqnarray}\label{rho1rho2}
	\frac{1}{r_1} = \frac{1}{r} \sum_{k=0}^{+\infty} \left( 1 + k\,X_1\,\delta X\,\frac{v_{12}^2}{2 c^2} \right) \left( X_2 \frac{r_{12}}{r} \right)^k P_k(n n_{12}) + \mathcal{O}(G,c^{-4}) \, ,
\end{eqnarray}
together with $1 \leftrightarrow 2$. Here the argument of the Legendre polynomial is $(n n_{12})$, the scalar product between $\mathbf{n}=\mathbf{x}/r$ and $\mathbf{n}_{12}=\mathbf{y}_{12}/r_{12}$. Notice that the term proportional to $G$ in Eq.~\eref{rel_pos} has consistently been neglected here. In addition, we see from Eqs.~\eref{metric2PN00}--\eref{metric2PNij} that the scalar products $(n_A v_A)$ are only needed with Newtonian accuracy. Making use of \eref{rel_pos}--\eref{rho1rho2}, we get 
\begin{eqnarray}\label{nAvA}
	\fl (n_1 v_1) = X_2 \left[ (n v_{12}) - X_2 \frac{r_{12}}{r} (n_{12} v_{12}) \right]  \sum_{k=0}^{+\infty} \left( X_2 \frac{r_{12}}{r} \right)^k P_k(n n_{12}) + \mathcal{O}(G,c^{-2}) \, .
\end{eqnarray}
We have a similar expansion for the term $n_1^i n_1^j / r_1$ appearing in the spatial metric $g_{ij}^{\rm PN}$.

Combining those results, we obtain the CL expansion of the 2PN metric \eref{metric2PN00}--\eref{metric2PNij}. In the companion paper \cite{Le.al.09} we shall use this expansion to compute the gravitational recoil generated during the ringdown phase. The gravitational recoil dominantly results from the coupling between the $\ell=2$ and $\ell=3$ multipole moments \cite{Sc.al.08}, where $\ell$ is the azimuthal number of the Schwarzschild perturbation. Thus we need to expand the 2PN metric at least up to octupolar order, i.e. up to $k=3$. In fact it will turn out that it is necessary to push the CL expansion up to $k=5$ if we want to control all the terms which are of order $r_{12}^3$ in the multipolar coefficients of the black hole perturbation in the Regge-Wheeler gauge. We shall discuss this point further in Sec.~\ref{secIIIsubII}.

At the zero-th order in the CL expansion we evidently recover the Schwarzschild metric of a black hole with mass $M=m_1+m_2$. Thanks to our gauge transformation \eref{xi0}--\eref{xii} we find it to be directly given in usual Schwarzschild coordinates, namely
\numparts
	\begin{eqnarray}
		g_{00}^{\rm Schw} & = - 1 + \frac{2 G M}{c^2 r} \, , \label{metricSchw00} \\
		g_{0i}^{\rm Schw} & = 0\, , \\ 
		g_{ij}^{\rm Schw} & = \delta^{ij} + \frac{2G M}{c^2 r} n^{i}n^{j} + \mathcal{O}(G^2) \, , \label{metricSchwij}
	\end{eqnarray}
\endnumparts
where $\mathbf{n} = \mathbf{x}/r$ is the unit vector in the direction of the field point $\mathbf{x}$. As expected, we find that the Schwarzschild metric is exactly recovered in the limit $\nu \rightarrow 0$, i.e. if one of the masses vanishes. Thus, we are now in a position to identify the 2PN metric expanded in the CL approximation with a perturbed Schwarzschild black hole with mass $M$, namely
\begin{equation}\label{PNsch}
	g_{\mu \nu}^{\rm PN} = g_{\mu \nu}^{\rm Schw} + h_{\mu \nu} \,.
\end{equation}
We find that the metric perturbation $h_{\mu\nu}$ expanded up to octupolar order in the CL approximation reads explicitly as
\numparts
	\begin{eqnarray}
		\fl h_{00} = \nu \frac{G M}{c^2 r} \frac{r_{12}^2}{r^2} \bigl[ 3 (n n_{12})^2 - 1 \bigr] - \nu \frac{G \delta M}{c^2 r} \frac{r_{12}^3}{r^3} (n n_{12}) \bigl[ 5 (n n_{12})^2 - 3 \bigr] + 3 \nu \frac{G M}{c^4 r} \bigl[ 2 v_{12}^2 - (n v_{12})^2 \bigr] \nonumber \\- \nu \frac{G \delta M}{c^4 r} \frac{r_{12}}{r} \biggl( 6 (n v_{12}) (n_{12} v_{12}) + (n n_{12}) \bigl[ 5 v_{12}^2 - 9 (n v_{12})^2 \bigr] \biggr) \nonumber \\ + 3 \nu (1 - 3 \nu) \frac{G M}{c^4 r} \frac{r_{12}^2}{r^2} \biggl( v_{12}^2 \bigl[ 3 (n n_{12})^2 - 1 \bigr] - \frac{3}{2} (n v_{12})^2 \bigl[ 5 (n n_{12})^2 - 1 \bigr] \nonumber \\ \qquad\qquad\qquad\qquad\qquad - (n_{12} v_{12})^2 + 6 (n_{12} v_{12}) (n v_{12}) (n n_{12}) \biggr) \nonumber \\ - \frac{3}{2} \nu (2 - 5 \nu) \frac{G \delta M}{c^4 r} \frac{r_{12}^3}{r^3} v_{12}^2 \bigl[ 5 (n n_{12})^2 - 3 \bigr] \nonumber \\ + 9 \nu (1 - 2 \nu) \frac{G \delta M}{c^4 r} \frac{r_{12}^3}{r^3} \biggl( (n n_{12}) (n_{12} v_{12})^2 + \frac{5}{6} (n n_{12}) (n v_{12})^2 \bigl[ 7 (n n_{12})^2 - 3 \bigr] \nonumber \\ \qquad\qquad\qquad- (n v_{12}) (n_{12} v_{12}) \bigl[ 5 (n n_{12})^2 - 1 \bigr] \biggr) + \mathcal{O}(G^2,c^{-6},r_{12}^4) \, , \label{metricPNCL00} \\
%%%%%%%%%%%%%%%%%%%%%%%%%%%%%%%%%%%%%%%%%%%%%%%%%%%%%%%%%%%%%%%
		\fl h_{0i} = - 4 \nu \frac{G M}{c^3 r} \frac{r_{12}}{r} (n n_{12}) v_{12}^i + 2 \nu \frac{G \delta M}{c^3 r} \frac{r_{12}^2}{r^2} \bigl[ 3 (n n_{12})^2 - 1 \bigr] v_{12}^i \nonumber \\ - 2 \nu (1 - 3 \nu) \frac{G M}{c^3 r} \frac{r_{12}^3}{r^3} (n n_{12}) \bigl[ 5 (n n_{12})^2 - 3 \bigr] v_{12}^i + 2 \nu \frac{G \delta M}{c^5 r} \bigl[ v_{12}^2 - (n v_{12})^2 \bigr] v_{12}^i \nonumber \\ - 2 \nu ( 1 - 3 \nu) \frac{G M}{c^5 r} \frac{r_{12}}{r} \biggl( 2 (n v_{12}) (n_{12} v_{12}) + (n n_{12}) \bigl[ 2 v_{12}^2 - 3 (n v_{12})^2 \bigr] \biggr) v_{12}^i \nonumber \\ + \nu ( 1 - 2 \nu ) \frac{G \delta M}{c^5 r} \frac{r_{12}^2}{r^2} \biggl( 2 (n_{12} v_{12}) \bigl[ 6 (n v_{12}) (n n_{12}) - (n_{12} v_{12}) \bigr] \nonumber \\ \qquad\qquad\qquad\qquad	- 3 (n v_{12})^2 \bigl[ 5 (n n_{12})^2 - 1 \big] \biggr) v_{12}^i \nonumber \\ + \nu ( 2 - 7 \nu ) \frac{G \delta M}{c^5 r} \frac{r_{12}^2}{r^2} v_{12}^2 \bigl[ 3 (n n_{12})^2 - 1 \bigr] v_{12}^i \nonumber \\ - 2 \nu ( 1 - 7 \nu + 13 \nu^2 ) \frac{G M}{c^5 r} \frac{r_{12}^3}{r^3} v_{12}^2 \bigl[ 5 (n n_{12})^2 - 3 \bigr] v_{12}^i \nonumber \\ + \nu ( 1 - 5 \nu + 5 \nu^2 ) \frac{G M}{c^5 r} \frac{r_{12}^3}{r^3} \biggl( 6 (n n_{12}) (n_{12} v_{12})^2 + 5 (n n_{12}) (n v_{12})^2 \bigl[ 7 (n n_{12})^2 - 3 \bigr] \nonumber \\ \qquad\qquad\quad - 6 (n v_{12}) (n_{12} v_{12}) \bigl[ 5 (n n_{12})^2 - 1 \bigr] \biggr) v_{12}^i + \mathcal{O}(G^2,c^{-6},r_{12}^4) \, , \label{metricPNCL0i} \\ 
%%%%%%%%%%%%%%%%%%%%%%%%%%%%%%%%%%%%%%%%%%%%%%%%%%%%%%%%%%%%%%%%%%%
		\fl h_{ij} = 2 \nu \frac{G M}{c^2 r} \frac{r_{12}^2}{r^2} \biggl( n_{12}^i n_{12}^j - 6 (n n_{12}) \, n_{12}^{(i} n^{j)} + \frac{3}{2} \bigl[ 5 (n n_{12})^2 - 1 \bigr] n^i n^j \biggr) \nonumber \\ - 6 \nu \frac{G \delta M}{c^2 r} \frac{r_{12}^3}{r^3} \biggl( (n n_{12}) \, n_{12}^i n_{12}^j - \bigl[ 5 (n n_{12})^2 - 1 \bigr] n_{12}^{(i} n^{j)} \nonumber \\ \qquad\qquad\qquad + \frac{5}{6} (n n_{12}) \bigl[ 7 (n n_{12})^2 - 3 \bigr] n^i n^j \biggr) \nonumber \\ + \nu \frac{G M}{c^4 r} \bigl[ 4 v_{12}^i v_{12}^j - (n v_{12})^2 \delta^{ij} \bigr] \nonumber \\ - \nu \frac{G \delta M}{c^4 r} \frac{r_{12}}{r} \biggl( 2 (n v_{12}) (n_{12} v_{12}) \delta^{ij} + 2 v_{12}^2 n_{12}^{(i} n^{j)} \nonumber \\ \qquad\qquad\qquad - 3 (n n_{12}) \bigl[ (n v_{12})^2 \delta^{ij} + v_{12}^2 n^i n^j \bigr] + 4 (n n_{12}) v_{12}^i v_{12}^j \biggr) \nonumber \\ + \nu (1 - 3 \nu) \frac{G M}{c^4 r} \frac{r_{12}^2}{r^2} \biggl( 2 \bigl[ 3 (n n_{12})^2 - 1 \bigr] v_{12}^i v_{12}^j - \frac{3}{2} (n v_{12})^2 \bigl[ 5 (n n_{12})^2 - 1 \bigr] \delta^{ij} \nonumber \\ \qquad\qquad\qquad\qquad\quad + (n_{12} v_{12}) \bigl[ 6 (n n_{12}) (n v_{12}) - (n_{12} v_{12} ) \bigr] \delta^{ij} \biggr) \nonumber \\ + \nu (1 - 2 \nu) \frac{G \delta M}{c^4 r} \frac{r_{12}^3}{r^3} \biggl( - 2 (n n_{12}) \bigl[ 5 (n n_{12})^2 - 3 \bigr] v_{12}^i v_{12}^j \nonumber \\ \qquad\qquad\qquad + \frac{5}{2} (n n_{12}) (n v_{12})^2 \bigl[ 7 (n n_{12})^2 - 3 \bigr] \delta^{ij} \nonumber \\ \qquad\qquad\qquad + 3 (n n_{12}) (n_{12} v_{12})^2 \delta^{ij} - 3 (n v_{12}) (n_{12} v_{12}) \bigl[ 5 (n n_{12})^2 - 1 \bigr] \delta^{ij} \nonumber \\ \qquad\qquad\qquad + 9 v_{12}^2 (n n_{12}) n_{12}^i n_{12}^j - 9 v_{12}^2 \bigl[ 5 (n n_{12})^2 - 1 \bigr] n_{12}^{(i} n^{j)} \nonumber \\ \qquad\qquad\qquad + \frac{15}{2} v_{12}^2 (n n_{12}) \bigl[ 7 (n n_{12})^2 - 3 \bigr] n^i n^j \biggr) + \mathcal{O}(G^2,c^{-6},r_{12}^4) \, . \label{metricPNCLij}
	\end{eqnarray}
\endnumparts
Even though we have performed the CL expansion including all terms up to order $r_{12}^5$, we only give in \eref{metricPNCL00}--\eref{metricPNCLij} the result up to octupolar order because of the proliferation of terms at higher orders. But we further stress that those terms proportional to $r_{12}^4$ and $r_{12}^5$ in Eqs.~\eref{metricPNCL00}--\eref{metricPNCLij} are fully under control (in an algebraic computer program), and were needed and used to get the final results given by Eqs.~\eref{polar1}--\eref{polar2} and \eref{axial1}--\eref{axial2} below.

If we come back for a moment to the three dimensionless scales $\varepsilon_{\rm PM}$, $\varepsilon_{\rm PN}$ and $\varepsilon_{\rm CL}$ defined in the Introduction, we can check that indeed the metric perturbation $h_{\mu\nu}$ admits the general structure given by Eq.~\eref{double_exp}, in which the dimensionless coefficients $h_{\mu\nu}^{(n,k)}$ are only functions of angles and mass ratios.

Although the identification \eref{PNsch} we are making is mathematically cristal clear, we recall that its physical justification is not completely straightforward. We invoke the theory of the matching of asymptotic series, but use it in a \textit{formal} way, since, as commented in the Introduction, the overlapping region between the domains of validity of the PN and CL expansions does not exist. Physically, we also rely on the fact that the merger as observed in numerical simulations lasts a very short time, which makes us feeling that the physics is essentially ``conserved'' when going from a PN description of the system to a perturbation of the final black hole. In addition, the PN approximation has proved to be very powerful in several past studies, with a domain of validity which often turned out to be larger than the one expected from elementary estimates (see \cite{Bl.al.09} for a recent example in the extreme mass ratio regime). Here we are assuming a rather extreme extension of the domain of validity of the PN expansion --- one for which $r_{12} \gtrsim G M/c^2$, corresponding to the ultra-relativistic limit $\varepsilon_{\rm PN} \lesssim 1$. Nevertheless we shall find below and in the application \cite{Le.al.09} that the PN approximation performs well.

In the next Section we shall describe our perturbation using the usual black hole perturbation formalism; for this we transform the coordinates from Cartesian $x^\mu\equiv\{c t,x,y,z\}$ to spherical $x^\alpha\equiv\{c t,r,\theta,\varphi\}$. This is appropriate for the spherically symmetric background, and the spherical coordinates are identified with the Schwarzschild(-Droste) coordinate system. Thus,
\begin{equation}
	h_{\alpha\beta}(x^\gamma) = \frac{\partial x^\mu}{\partial x^\alpha} \,\frac{\partial x^\nu}{\partial x^\beta} \,h_{\mu\nu}(x^\rho)\, .
\end{equation}
We then write the scalar products $(n n_{12})$ and $(n v_{12})$ in terms of them. Our conventions regarding orientations and various angles are explained in Fig.~\ref{fig_binary}. The unit vector $\mathbf{n}=\mathbf{x}/r$ pointing from the center of mass to the field point, and the unit direction $\mathbf{n}_{12} = \mathbf{y}_{12}/r_{12}$ from body $2$ to body $1$ read
\begin{eqnarray}
	\mathbf{n} & = \left( \sin{\theta} \cos{\varphi}, \sin{\theta} \sin{\varphi}, \cos{\theta} \right) , \\
	\mathbf{n}_{12} & = \left( \cos{\beta}, \sin{\beta}, 0 \right) .
\end{eqnarray}
Note our unconventional notation for the orbital phase angle $\beta$. For a generic non-circular orbit, we have (see Fig.~\ref{fig_binary})
\numparts
	\begin{eqnarray}
		(n n_{12}) &= \sin{\theta} \cos{(\varphi - \beta)} \, , \\
		(n v_{12}) &= \sin{\theta} \, \bigl[ \dot{r}_{12} \cos{(\varphi - \beta)} + r_{12} \omega_{12} \sin{(\varphi - \beta)} \bigr] \, .
	\end{eqnarray}
\endnumparts
Here the relative angular velocity is $\omega_{12} \equiv \dot{\beta}$ with $\beta$ being the orbital phase, and $\dot{r}_{12} \equiv (n_{12} v_{12})$ is the inspiral rate, where a dot stands for a derivative with respect to coordinate time $t$.  Finally we find that all the components of the perturbation $h_{\alpha \beta}$ of the Schwarzschild metric $g_{\alpha \beta}^{\rm Schw}$ (both written in spherical coordinates $x^\alpha=\{c t,r,\theta,\varphi\}$) are given as explicit functions of the spherical coordinates $\{r,\theta,\varphi\}$, and depend on time $t$ through the orbital parameters $\beta$, $\omega_{12}$, $r_{12}$ and $\dot{r}_{12}$.

\section{The 2PN metric in Regge-Wheeler-Zerilli formalism}
\label{secIII}

\subsection{Multipole decomposition of a Schwarzschild perturbation}
\label{secIIIsubI}

We briefly remind (see e.g. \cite{No.99,MaPo.05,NaRe.05,Gl.al.00} for more details) the usual decomposition into multipoles of a first-order perturbation of a Schwarzschild black hole of mass $M$. As usual, we write the perturbation $h_{\alpha \beta}$ as the sum of two kinds of perturbations,
\begin{equation}
	h_{\alpha \beta} = h_{\alpha \beta}^{({\rm e})} + h_{\alpha \beta}^{({\rm o})}\, ,
\end{equation}	
where the even-parity perturbation $h_{\alpha \beta}^{({\rm e})}$ essentially describes a perturbation along an (arbitrary) axis of the spherically symmetric Schwarzschild background, and where the odd-parity perturbation $h_{\alpha \beta}^{({\rm o})}$ essentially describes a perturbation around that axis.\footnote{The even-parity perturbation is often called the ``polar'' perturbation, while the odd-parity one is called the ``axial'' perturbation. (From now on we pose $G = c = 1$.)} Both perturbations are expanded with respect to a set of 10 tensorial spherical harmonics (cf.~\ref{appA}).

Following Regge and Wheeler's \cite{ReWh.57} conventions, the even-parity perturbation multipole decomposition reads \cite{Gl.al.00} 
\numparts
	\begin{eqnarray}
		h_{00}^{({\rm e})} & = \left( 1 - \frac{2M}{r} \right) \sum_{\ell,m} H_0^{\ell,m} Y_{\ell,m} \, , \label{polardecomp1} \\
		h_{0r}^{({\rm e})} & = \sum_{\ell,m} H_1^{\ell,m} Y_{\ell,m} \, , \\
		h_{0\theta}^{({\rm e})} & = \sum_{\ell,m} h_0^{\ell,m} \partial_\theta Y_{\ell,m} \, , \\
		h_{0\varphi}^{({\rm e})} & = \sum_{\ell,m} h_0^{\ell,m} \partial_\varphi Y_{\ell,m} \, , \\
		h_{rr}^{({\rm e})} & = \left( 1 - \frac{2M}{r} \right)^{-1} \sum_{\ell,m} H_2^{\ell,m} Y_{\ell,m} \, , \\
		h_{r\theta}^{({\rm e})} & = \sum_{\ell,m} h_1^{\ell,m} \partial_\theta Y_{\ell,m} \, , \\
		h_{r\varphi}^{({\rm e})} & = \sum_{\ell,m} h_1^{\ell,m} \partial_\varphi Y_{\ell,m} \, , \\
		h_{\theta \theta}^{({\rm e})} & = r^2 \sum_{\ell,m} \left( K^{\ell,m} + G^{\ell,m} \partial^2_\theta \right) Y_{\ell,m} \, , \\
		h_{\theta \varphi}^{({\rm e})} & = r^2 \sum_{\ell,m} G^{\ell,m} \left( \partial^2_{\theta \varphi} - \cot{\theta} \, \partial_\varphi \right) Y_{\ell,m} \, , \\
		h_{\varphi \varphi}^{({\rm e})} & = r^2 \sum_{\ell,m} \left[ K^{\ell,m} \sin^2{\theta} + G^{\ell,m} \left( \partial_\varphi^2 + \sin{\theta} \cos{\theta} \, \partial_\theta \right) \right] Y_{\ell,m} \, , \label{polardecomp2}
	\end{eqnarray}
\endnumparts
where the summations over the integers $\ell$ and $m$ range from 2 to infinity, and from $-\ell$ to $\ell$ respectively. Note that the low multipoles $\ell = 0$ and $\ell = 1$ correspond to the non-radiating pieces of the perturbation $h_{\alpha \beta}$, and are not relevant to gravitational waves. For example, a monopolar perturbation ($\ell = 0$) would correspond to an infinitesimal shift of the black hole mass. (See e.g. \cite{Ze1.70,MaPo.05} for more details.) Similarly, the multipole decomposition of the odd-parity perturbation is
\numparts
	\begin{eqnarray}
		h_{0 \theta}^{({\rm o})} & = - \sum_{\ell,m} k_0^{\ell,m} \frac{\partial_\varphi Y_{\ell,m}}{\sin{\theta}} \, , \label{axialdecomp1} \\
		h_{0 \varphi}^{({\rm o})} & = \sum_{\ell,m} k_0^{\ell,m} \sin{\theta} \, \partial_\theta Y_{\ell,m} \, , \\
		h_{r \theta}^{({\rm o})} & = - \sum_{\ell,m} k_1^{\ell,m} \frac{\partial_\varphi Y_{\ell,m}}{\sin{\theta}} \, , \\
		h_{r \varphi}^{({\rm o})} & = \sum_{\ell,m} k_1^{\ell,m} \sin{\theta} \, \partial_\theta Y_{\ell,m} \, , \\
		h_{\theta \theta}^{({\rm o})} & = \sum_{\ell,m} k_2^{\ell,m} \frac{1}{\sin{\theta}} \left( \partial_\theta - \cot{\theta} \right) \partial_\varphi Y_{\ell,m} \, , \\
		h_{\theta \varphi}^{({\rm o})} & = \frac{1}{2} \sum_{\ell,m} k_2^{\ell,m} \frac{1}{\sin{\theta}} \left( \partial^2_\varphi + \cos{\theta} \sin{\theta} \, \partial_\theta - \sin^2{\theta} \, \partial^2_\theta \right) Y_{\ell,m} \, , \\
		h_{\varphi \varphi}^{({\rm o})} & = - \sum_{\ell,m} k_2^{\ell,m} \sin{\theta} \left( \partial_\theta - \cot{\theta} \right) \partial_\varphi Y_{\ell,m} \, . \label{axialdecomp2}
	\end{eqnarray}
\endnumparts
Our convention for the scalar spherical harmonics $Y_{\ell,m}$ is given in Eq.~\eref{Ylm} of \ref{appA}. In Eqs.~\eref{polardecomp1}--\eref{polardecomp2} and \eref{axialdecomp1}--\eref{axialdecomp2} all the multipolar coefficients $H_0^{\ell,m}$, $H_1^{\ell,m}$, $H_2^{\ell,m}$, $K^{\ell,m}$, $G^{\ell,m}$, $h_0^{\ell,m}$, $h_1^{\ell,m}$, $k_0^{\ell,m}$, $k_1^{\ell,m}$ and $k_2^{\ell,m}$ are functions of $\{ t,r \}$ in Schwarzschild coordinates, and are defined in an arbitrary perturbative gauge.

\subsection{Computation of the multipole contributions}
\label{secIIIsubII}

Given the metric perturbation $h_{\alpha \beta}$ obtained from the CL approximation in the previous Section, we can compute all the coefficients $H_0^{\ell,m},H_1^{\ell,m},\cdots,k_2^{\ell,m}$. After some calculations consisting mostly of projections over the Zerilli-Mathews tensor spherical harmonics (using their orthonormality properties recalled in \ref{appA}), we obtain the even and odd multipolar coefficients in a particular gauge, which follows from the choice of gauge made in Eq.~\eref{xi0}--\eref{xii}. We shall however change to the Regge-Wheeler gauge, where the multipolar coefficients $G^{\ell,m}$, $h_0^{\ell,m}$, $h_1^{\ell,m}$ and $k_2^{\ell,m}$ vanish; this makes the expressions of the Regge-Wheeler and Zerilli master functions much simpler [see Eqs.~\eref{Z_master}--\eref{RW_master} below]. Note the difference between the choice of gauge \eref{xi0}--\eref{xii}, which was made for the PN metric before its CL expansion, and a choice of gauge within black hole perturbation theory, once the PN metric is in CL form. We transform the results to the Regge-Wheeler gauge by making the substitutions (see e.g. \cite{Gl.al.00,Na.al.03} for general expressions):
\numparts
	\begin{eqnarray}
		\fl H_0^{\ell,m} & \longrightarrow \widetilde{H}_0^{\ell,m} & = H_0^{\ell,m} - 2 \partial_t h_0^{\ell,m} + r^2 \partial^2_t G^{\ell,m} + \mathcal{O}(G^2) \, , \label{H0RW} \\
		\fl H_1^{\ell,m} & \longrightarrow \widetilde{H}_1^{\ell,m} & = H_1^{\ell,m} - \partial_r h_0^{\ell,m} - \partial_t h_1^{\ell,m} + r \partial_t G^{\ell,m} + r^2 \partial^2_{tr} G^{\ell,m} + \mathcal{O}(G^2) \, , \\
		\fl H_2^{\ell,m} & \longrightarrow \widetilde{H}_2^{\ell,m} & = H_2^{\ell,m} - 2 \partial_r h_1^{\ell,m} + 2 r \partial_r G^{\ell,m} + r^2 \partial^2_r G^{\ell,m} + \mathcal{O}(G^2) \, , \\
		\fl K^{\ell,m} & \longrightarrow \widetilde{K}^{\ell,m} & = K^{\ell,m} + r \partial_r G^{\ell,m} - 2 h_1^{\ell,m} / r + \mathcal{O}(G^2) \, , \\
		\fl k_0^{\ell,m} & \longrightarrow \widetilde{k}_0^{\ell,m} & = k_0^{\ell,m} + \partial_t k_2^{\ell,m} / 2 \, , \\
		\fl k_1^{\ell,m} & \longrightarrow \widetilde{k}_1^{\ell,m} & = k_1^{\ell,m} + \partial_r k_2^{\ell,m} / 2 - k_2^{\ell,m} / r \, . \label{k1RW}
	\end{eqnarray}
\endnumparts
We can now understand why it was necessary to expand the 2PN metric so as to include terms of order $r_{12}^4$ and $r_{12}^5$ in the initial perturbation \eref{metricPNCL00}--\eref{metricPNCLij}. In the gauge transformation \eref{H0RW}--\eref{k1RW}, the partial time derivatives of the multipolar coefficients in the initial gauge yield lower order powers of $r_{12}$ in the expression of the new multipolar coefficients in Regge-Wheeler's gauge. For example, from \eref{H0RW} we observe that terms like $G^{3,\pm3} \sim r_{12}^5/c^2$ in the initial gauge produce terms like $c^{-2}\partial^2_t G^{3,\pm3} \sim \dot{r}_{12}^2 r_{12}^3 / c^4$ in the multipolar coefficients $\widetilde{H}_0^{3,\pm3}$ in the Regge-Wheeler gauge. Such contributions of order $r_{12}^4$ and $r_{12}^5$ in the multipolar coefficients describing the perturbation in the initial gauge need to be consistently included in order to control all terms of order $r_{12}^3$ in the multipolar coefficients describing the final perturbation in the Regge-Wheeler gauge.

As expected, we find that the simplest of the Einstein field equations in vacuum in the Regge-Wheeler gauge is satisfied [up to terms $\mathcal{O}(G^2,c^{-6},r_{12}^4)$], namely
\begin{equation}\label{H0H2}
	\widetilde{H}_0^{\ell,m} = \widetilde{H}_2^{\ell,m} \equiv \widetilde{H}^{\ell,m} \, .
\end{equation}
Finally, we give below all the non-zero multipolar coefficients in the Regge-Wheeler gauge for all $(\ell,m)$ up to $\ell = 3$. All equations below are valid modulo remainder terms $\mathcal{O}(G^2,c^{-6},r_{12}^4) $. For the even-parity perturbation,
\numparts
	\begin{eqnarray}
		\fl \widetilde{H}^{2,0} = - 2 \sqrt{\frac{\pi}{5}} \, \nu \, \frac{M}{r} \biggl\{ \frac{r_{12}^2}{r^2} \biggl[ 1 + \bigl( 1 - 3 \nu \bigr) \biggl( \frac{9}{14} \dot{r}_{12}^2 + \frac{17}{14} r_{12}^2 \omega_{12}^2 \biggr) \biggr] + v_{12}^2 \biggr\} \, , \label{polar1} \\
		\fl \widetilde{H}^{2,\pm 2} = \sqrt{\frac{6 \pi}{5}} \, \nu \, \frac{M}{r} \biggl\{ \frac{r_{12}^2}{r^2} \biggl[ 1 + \bigl( 1 - 3 \nu \bigr) \biggl( \frac{9}{14} \dot{r}_{12}^2 \mp \frac{10}{21} \rmi \, r_{12} \omega_{12} \dot{r}_{12} + \frac{1}{6} r_{12}^2 \omega_{12}^2 \biggr) \biggr] \nonumber \\ + \bigl( \dot{r}_{12} \mp \rmi \, r_{12} \omega_{12} \bigr)^2 \biggr\} e^{\mp 2 \rmi \beta} \, , \label{H22} \\
		\fl \widetilde{H}^{3,\pm 1} = \mp \sqrt{\frac{3 \pi}{7}} \, \nu \, \frac{\delta M r_{12}}{r^2} \biggl\{ \frac{r_{12}^2}{r^2} \biggl[ 1 + \frac{5}{6} \biggl( 1 - \frac{19}{5} \nu \biggr) \dot{r}_{12}^2 \mp \frac{1}{3} \bigl( 1 - 2 \nu \bigr) \, \rmi \, r_{12} \omega_{12} \dot{r}_{12} \nonumber \\ + \frac{7}{6} \biggl( 1 - \frac{23}{7} \nu \biggr) r_{12}^2 \omega_{12}^2 \biggr] + \biggl( \dot{r}_{12}^2 \mp \frac{2}{3} \, \rmi \, r_{12} \omega_{12} \dot{r}_{12} + \frac{1}{3} r_{12}^2 \omega_{12}^2 \biggr) \biggr\} e^{\mp \rmi \beta} \, , \\
		\fl \widetilde{H}^{3,\pm 3} = \pm \sqrt{\frac{5 \pi}{7}} \, \nu \, \frac{\delta M r_{12}}{r^2} \biggl\{ \frac{r_{12}^2}{r^2} \biggl[ 1 + \frac{5}{6} \biggl( 1 - \frac{19}{5} \nu \biggr) \dot{r}_{12}^2 \mp \bigl( 1 - 2 \nu \bigr) \, \rmi \, r_{12} \omega_{12} \dot{r}_{12} \nonumber \\ - \frac{1}{6} \bigl( 1 + 7 \nu \bigr) r_{12}^2 \omega_{12}^2 \biggr] + (\dot{r}_{12} \mp \rmi \, r_{12} \omega_{12})^2 \biggr\} e^{\mp 3 \rmi \beta} \, , \\
		\fl \widetilde{K}^{2,0} = - 2 \sqrt{\frac{\pi}{5}} \, \nu \, \frac{M}{r} \biggl\{ \frac{r_{12}^2}{r^2} \biggl[ 1 + \bigl( 1 - 3 \nu \bigr) \biggl( \frac{9}{14} \dot{r}_{12}^2 + \frac{17}{14} r_{12}^2 \omega_{12}^2 \biggr) \biggr] - v_{12}^2 \biggr\} \, , \label{K20} \\
		\fl \widetilde{K}^{2,\pm 2}  = \sqrt{\frac{6 \pi}{5}} \, \nu \, \frac{M}{r} \biggl\{ \frac{r_{12}^2}{r^2} \biggl[ 1 + \bigl( 1 - 3 \nu \bigr) \biggl( \frac{9}{14} \dot{r}_{12}^2 \mp \frac{10}{21} \rmi \, r_{12} \omega_{12} \dot{r}_{12} + \frac{1}{6} r_{12}^2 \omega_{12}^2 \biggr) \biggr] \nonumber \\ - \bigl( \dot{r}_{12} \mp \rmi \, r_{12} \omega_{12} \bigr)^2 \biggr\} e^{\mp 2 \rmi \beta} \, , \label{K22} \\
		\fl \widetilde{K}^{3,\pm 1} = \mp \sqrt{\frac{3 \pi}{7}} \, \nu \, \frac{\delta M r_{12}}{r^2} \biggl\{ \frac{r_{12}^2}{r^2} \biggl[ 1 + \frac{5}{6} \biggl( 1 - \frac{19}{5} \nu \biggr) \dot{r}_{12}^2 \mp \frac{1}{3} \bigl( 1 - 2 \nu \bigr) \, \rmi \, r_{12} \omega_{12} \dot{r}_{12} \nonumber \\ + \frac{7}{6} \biggl( 1 - \frac{23}{7} \nu \biggr) r_{12}^2 \omega_{12}^2 \biggr] - \biggl( \dot{r}_{12}^2 \mp \frac{2}{3} \, \rmi \, r_{12} \omega_{12} \dot{r}_{12} + \frac{1}{3} r_{12}^2 \omega_{12}^2 \biggr) \biggr\} e^{\mp \rmi \beta} \, , \\
		\fl \widetilde{K}^{3,\pm 3} = \pm \sqrt{\frac{5 \pi}{7}} \, \nu \, \frac{\delta M r_{12}}{r^2} \biggl\{ \frac{r_{12}^2}{r^2} \biggl[ 1 + \frac{5}{6} \biggl( 1 - \frac{19}{5} \nu \biggr) \dot{r}_{12}^2 \mp \bigl( 1 - 2 \nu \bigr) \, \rmi \, r_{12} \omega_{12} \dot{r}_{12} \nonumber \\ - \frac{1}{6} \bigl( 1 + 7 \nu \bigr) r_{12}^2 \omega_{12}^2 \biggr] - (\dot{r}_{12} \mp \rmi \, r_{12} \omega_{12})^2 \biggr\} e^{\mp 3 \rmi \beta} \, , \\
		\fl \widetilde{H}_1^{2,0} = 4 \sqrt{\frac{\pi}{5}} \, \nu \, \frac{M r_{12}}{r^2} \, \dot{r}_{12} \biggl[ 1 + \frac{9}{14} \bigl( 1 - 3 \nu \bigr) v_{12}^2 \biggr] \, , \\
		\fl \widetilde{H}_1^{2,\pm 2}  = - 2 \sqrt{\frac{6 \pi}{5}} \, \nu \, \frac{M r_{12}}{r^2} \, \bigl( \dot{r}_{12} \mp \rmi \, r_{12} \omega_{12} \bigr) \biggl[ 1 + \bigl( 1 - 3 \nu \bigr) \biggl( \frac{9}{14} \dot{r}_{12}^2 \mp \frac{5}{21} \, \rmi \, r_{12} \omega_{12} \dot{r}_{12} \nonumber \\ + \frac{17}{42} r_{12}^2 \omega_{12}^2 \biggr) \biggr] e^{\mp 2 \rmi \beta} \, , \\
		\fl \widetilde{H}_1^{3,\pm 1}  = \pm 2 \sqrt{\frac{3 \pi}{7}} \, \nu \, \frac{\delta M r_{12}^2}{r^3} \biggl[ \dot{r}_{12} \mp \frac{1}{3} \, \rmi \, r_{12} \omega_{12} + v_{12}^2 \biggl\{ \frac{5}{6} \biggl( 1 - \frac{19}{5} \nu \biggr) \dot{r}_{12} \nonumber \\ \mp \frac{1}{2} \bigl( 1 - 3 \nu \bigr) \rmi \, r_{12} \omega_{12} \biggr\} \biggr] e^{\mp \rmi \beta} \, , \\
		\fl \widetilde{H}_1^{3,\pm 3}  = \mp 2 \sqrt{\frac{5 \pi}{7}} \, \nu \, \frac{\delta M r_{12}^2}{r^3}\bigl( \dot{r}_{12} \mp \rmi \, r_{12} \omega_{12} \bigr) \biggl[ 1 + \frac{5}{6} \biggl( 1 - \frac{19}{5} \nu \biggr) \dot{r}_{12}^2 \nonumber \\ \mp \frac{2}{3} \bigl( 1 - 2 \nu \bigr) \rmi \, r_{12} \omega_{12} \dot{r}_{12} + \frac{1}{6} \bigl( 1 - 11 \nu \bigr) r_{12}^2 \omega_{12}^2 \biggr) \biggr] e^{\mp 3 \rmi \beta} \, , \label{polar2}
	\end{eqnarray}
\endnumparts
where we made use of the relation $v_{12}^2 = \dot{r}_{12}^2 + r_{12}^2 \omega_{12}^2$ when convenient. Similarly, the non-zero multipolar coefficients for the odd-parity perturbation are
\numparts
	\begin{eqnarray}
		\fl \widetilde{k}_0^{2,\pm 1} = \pm 2 \sqrt{\frac{2 \pi}{15}} \, \nu \, \frac{\delta M r_{12}^2}{r^2} \, r_{12} \omega_{12} \biggl[ 1 + \frac{9}{14} \biggl( 1 - \frac{13}{3} \nu \biggr) \dot{r}_{12}^2 \mp \frac{5}{28} \bigl( 1 - 2 \nu \bigr) \rmi \, r_{12} \omega_{12} \dot{r}_{12} \nonumber \\ + \frac{13}{28} \biggl( 1 - \frac{68}{13} \nu \biggr) r_{12}^2 \omega_{12}^2 \biggr] e^{\mp \rmi \beta} \, , \label{axial1} \\
		\fl \widetilde{k}_0^{3,0} = - \sqrt{\frac{\pi}{7}} \, \nu \, \frac{M r_{12}^3}{r^3} \, r_{12} \omega_{12} \biggl[ \bigl( 1 - 3 \nu \bigr) + \biggl( \frac{5}{6} - \frac{37}{6} \nu + \frac{73}{6} \nu^2 \biggr) \dot{r}_{12}^2 \nonumber \\ + \biggl( \frac{11}{18} - \frac{91}{18} \nu + \frac{199}{18} \nu^2 \biggr) r_{12}^2 \omega_{12}^2 \biggr] \, , \\
		\fl \widetilde{k}_0^{3,\pm 2} = \frac{1}{3} \sqrt{\frac{15 \pi}{14}} \, \nu \, \frac{M r_{12}^3}{r^3} \, r_{12} \omega_{12} \biggl[ \bigl( 1 - 3 \nu \bigr) + \biggl( \frac{5}{6} - \frac{37}{6} \nu + \frac{73}{6} \nu^2 \biggr) \dot{r}_{12}^2 \nonumber \\ \mp \biggl( \frac{8}{15} - \frac{8}{3} \nu + \frac{8}{3} \nu^2 \biggr) \rmi \, r_{12} \omega_{12} \dot{r}_{12} + \biggl( \frac{3}{10} - \frac{7}{2} \nu + \frac{19}{2} \nu^2 \biggr) r_{12}^2 \omega_{12}^2 \biggr] e^{\mp 2 \rmi \beta} \, , \\
		\fl \widetilde{k}_1^{2,\pm 1} = \mp \sqrt{\frac{8 \pi}{15}} \, \nu \, \frac{\delta M r_{12}}{r} \bigl( \dot{r}_{12} \mp \rmi \, r_{12} \omega_{12} \bigr) r_{12} \omega_{12} \, e^{\mp \rmi \beta} \, , \\
		\fl \widetilde{k}_1^{3,0} = \sqrt{\frac{\pi}{7}} \, \nu \, \frac{M r_{12}^2}{r^2} \bigl( 1 - 3 \nu \bigr) r_{12} \omega_{12} \dot{r}_{12} \, , \\
		\fl \widetilde{k}_1^{3,\pm 2} = - \sqrt{\frac{5 \pi}{42}} \, \nu \, \frac{M r_{12}^2}{r^2} \bigl( 1 - 3 \nu \bigr) \bigl( \dot{r}_{12} \mp \rmi \, r_{12} \omega_{12} \bigr) r_{12} \omega_{12} \, e^{\mp 2 \rmi \beta} \, . \label{axial2}
	\end{eqnarray}
\endnumparts
One can check that any given multipolar coefficient $F^{\ell,m}$ in \eref{polar1}--\eref{polar2} and \eref{axial1}--\eref{axial2} satisfies the property $F^{\ell,-m} = (-1)^m \, \bar{F}^{\ell,m}$, where the overbar denotes the complex conjugation, consistently with the fact that the initial perturbation is real-valued. Furthermore, all the multipolar coefficients associated with the odd (or axial) perturbation vanish as expected in the zero angular momentum limit $\omega_{12} = 0$, which corresponds to purely radial infall. We shall consider such head-on collisions in Sec.~\ref{secV}.

\section{Verification of the Einstein field equations}\label{secIV}

As an important check of the previous results, we now verify that all the perturbative Einstein equations are satisfied. This requires the computation of the partial time derivatives of the multipolar coefficients \eref{polar1}--\eref{polar2} and \eref{axial1}--\eref{axial2}. Recall that a generic multipolar coefficient $F^{\ell,m}$ is function of the coordinate time $t$ through the orbital phase $\beta(t)$, the orbital frequency $\omega_{12}(t) = \dot{\beta}(t)$, the distance $r_{12}(t)$, and the inspiral rate $\dot{r}_{12}(t)$. Therefore, we have
\begin{equation}\label{dtF}
	\partial_t F^{\ell,m} = \omega_{12} \frac{\partial F^{\ell,m}}{\partial \beta} + \dot{\omega}_{12} \frac{\partial F^{\ell,m}}{\partial \omega_{12}} + \dot{r}_{12} \frac{\partial F^{\ell,m}}{\partial r_{12}} + \ddot{r}_{12} \frac{\partial F^{\ell,m}}{\partial \dot{r}_{12}} \, .
\end{equation}
The relative position, velocity, and acceleration of the two bodies can be expressed as
\numparts\label{frenet}
\begin{eqnarray}
	\mathbf{y}_{12} &= r_{12} \,\mathbf{n}_{12}\, , \\
	\mathbf{v}_{12} &= \dot r_{12} \,\mathbf{n}_{12} + r_{12} \,\omega_{12} \,\bm{\lambda}_{12}\, , \\
	\mathbf{a}_{12} &= (\ddot r_{12} - r_{12} \,\omega_{12}^2) \,\mathbf{n}_{12} + 
(r_{12} \,\dot\omega_{12} + 2 \dot r_{12} \,\omega_{12}) \,\bm{\lambda}_{12}\, ,
\end{eqnarray}\endnumparts
where we have introduced the Frenet frame ($\mathbf{n}_{12},\bm{\lambda}_{12}$) defined by $\bm{\lambda}_{12}=\hat{\mathbf{L}} \times \mathbf{n}_{12}$, with $\hat{\mathbf{L}}$ being the unit vector orthogonal to the orbital plane, and in the same direction as the orbital angular momentum. Since we are working at linear order in $G$, the acceleration $\mathbf{a}_{12}$ which is proportional to $G$ can be neglected here, and we have 
\numparts
	\begin{eqnarray}
		\ddot{r}_{12} &= \omega_{12}^2 r_{12} + \mathcal{O}(G) \, , \\
		\dot{\omega}_{12} &= - 2 \omega_{12} \frac{\dot{r}_{12}}{r_{12}} + \mathcal{O}(G) \, .
	\end{eqnarray}
\endnumparts
Introducing these expressions of $\ddot{r}_{12}$ and $\dot{\omega}_{12}$ into \eref{dtF}, and neglecting terms $\mathcal{O}(G^2,c^{-6},r_{12}^4)$, we find for the even perturbation the non-zero partial time derivatives
\numparts
	\begin{eqnarray}
		\fl \partial_t \widetilde{H}^{2,0} = - 4 \sqrt{\frac{\pi}{5}} \, \nu \, \frac{M r_{12}}{r^3} \, \dot{r}_{12} \biggl[ 1 + \frac{9}{14} \bigl( 1 - 3 \nu \bigr) v_{12}^2 \biggr] \, , \label{dtpolar1} \\
		\fl \partial_t \widetilde{H}^{2,\pm 2} = 2 \sqrt{\frac{6 \pi}{5}} \, \nu \, \frac{M r_{12}}{r^3} \bigl( \dot{r}_{12} \mp \rmi \, r_{12} \omega_{12} \bigr) \biggl[ 1 + \bigl( 1 - 3 \nu \bigr) \biggl( \frac{9}{14} \dot{r}_{12}^2 \mp \frac{5}{21} \rmi \, r_{12} \omega_{12} \dot{r}_{12} \nonumber \\ + \frac{17}{42} r_{12}^2 \omega_{12}^2 \biggr) \biggr] e^{\mp 2 \rmi \beta} \, , \\
		\fl \partial_t \widetilde{H}^{3,\pm 1} = \mp \sqrt{\frac{3 \pi}{7}} \, \nu \, \frac{\delta M}{r^2} \biggl\{ \frac{r_{12}^2}{r^2} \biggl[ \bigl( 3 \dot{r}_{12} \mp \rmi \, r_{12} \omega_{12} \bigr) + \frac{5}{2} \biggl( 1 - \frac{19}{5} \nu \biggr) v_{12}^2 \dot{r}_{12} \nonumber \\ \mp \frac{3}{2} \bigl( 1 - 3 \nu \bigr) v_{12}^2 \, \rmi \, r_{12} \omega_{12} \biggr] + v_{12}^2 \bigl( \dot{r}_{12} \mp \rmi \, r_{12} \omega_{12} \bigr) \biggr\} e^{\mp \rmi \beta} \, , \\
		\fl \partial_t \widetilde{H}^{3,\pm 3} = \pm \sqrt{\frac{5 \pi}{7}} \, \nu \, \frac{\delta M}{r^2} \bigl( \dot{r}_{12} \mp \rmi \, r_{12} \omega_{12} \bigr) \biggl\{ \frac{r_{12}^2}{r^2} \biggl[ 3 + \frac{5}{2} \biggl( 1 - \frac{19}{5} \nu \biggr) \dot{r}_{12}^2 \nonumber \\ \mp 2 \bigl( 1 - 2 \nu \bigr) \, \rmi \, r_{12} \omega_{12} \dot{r}_{12} + \frac{1}{2} \bigl( 1 - 11 \nu \bigr) r_{12}^2 \omega_{12}^2 \biggr] + (\dot{r}_{12} \mp \, \rmi \, r_{12} \omega_{12})^2 \biggr\} e^{\mp 3 \rmi \beta} \, , \\
		\fl \partial_t \widetilde{K}^{2,0} = - 4 \sqrt{\frac{\pi}{5}} \, \nu \, \frac{M r_{12}}{r^3} \, \dot{r}_{12} \biggl[ 1 + \frac{9}{14} \bigl( 1 - 3 \nu \bigr) v_{12}^2 \biggr] \, , \\
		\fl \partial_t \widetilde{K}^{2,\pm 2} = 2 \sqrt{\frac{6 \pi}{5}} \, \nu \, \frac{M r_{12}}{r^3} \bigl( \dot{r}_{12} \mp \rmi \, r_{12} \omega_{12} \bigr) \biggl[ 1 + \bigl( 1 - 3 \nu \bigr) \biggl( \frac{9}{14} \dot{r}_{12}^2 \mp \frac{5}{21} \rmi \, r_{12} \omega_{12} \dot{r}_{12} \nonumber \\ + \frac{17}{42} r_{12}^2 \omega_{12}^2 \biggr) \biggr] e^{\mp 2 \rmi \beta} \, , \\
		\fl \partial_t \widetilde{K}^{3,\pm 1} = \mp \sqrt{\frac{3 \pi}{7}} \, \nu \, \frac{\delta M}{r^2} \biggl\{ \frac{r_{12}^2}{r^2} \biggl[ \bigl( 3 \dot{r}_{12} \mp \rmi \, r_{12} \omega_{12} \bigr) + \frac{5}{2} \biggl( 1 - \frac{19}{5} \nu \biggr) v_{12}^2 \dot{r}_{12} \nonumber \\ \mp \frac{3}{2} \bigl( 1 - 3 \nu \bigr) v_{12}^2 \, \rmi \, r_{12} \omega_{12} \biggr] - v_{12}^2 \bigl( \dot{r}_{12} \mp \rmi \, r_{12} \omega_{12} \bigr) \biggr\} e^{\mp \rmi \beta} \, , \\
		\fl \partial_t \widetilde{K}^{3,\pm 3} = \pm \sqrt{\frac{5 \pi}{7}} \, \nu \, \frac{\delta M}{r^2} \bigl( \dot{r}_{12} \mp \rmi \, r_{12} \omega_{12} \bigr) \biggl\{ \frac{r_{12}^2}{r^2} \biggl[ 3 + \frac{5}{2} \biggl( 1 - \frac{19}{5} \nu \biggr) \dot{r}_{12}^2 \nonumber \\ \mp 2 \bigl( 1 - 2 \nu \bigr) \, \rmi \, r_{12} \omega_{12} \dot{r}_{12} + \frac{1}{2} \bigl( 1 - 11 \nu \bigr) r_{12}^2 \omega_{12}^2 \biggr] - (\dot{r}_{12} \mp \rmi \, r_{12} \omega_{12})^2 \biggr\} e^{\mp 3 \rmi \beta} \, , \\
		\fl \partial_t \widetilde{H}_1^{2,0} = 4 \sqrt{\frac{\pi}{5}} \, \nu \, \frac{M}{r^2} \, v_{12}^2 \biggl[ 1 + \frac{9}{14} \bigl( 1 - 3 \nu \bigr) v_{12}^2 \biggr] \, , \\
		\fl \partial_t \widetilde{H}_1^{2,\pm 2} = - 2 \sqrt{\frac{6 \pi}{5}} \, \nu \, \frac{M}{r^2} \, \bigl( \dot{r}_{12} \mp \rmi \, r_{12} \omega_{12} \bigr)^2 \biggl[ 1 + \frac{9}{14} \bigl( 1 - 3 \nu \bigr)  v_{12}^2 \biggr] e^{\mp 2 \rmi \beta} \, , \\
		\fl \partial_t \widetilde{H}_1^{3,\pm 1} = \pm 4 \sqrt{\frac{3 \pi}{7}} \, \nu \, \frac{\delta M r_{12}}{r^3} \bigl( \dot{r}_{12} \mp \rmi \, r_{12} \omega_{12} \bigr) \biggl[ \dot{r}_{12} \pm \frac{1}{3} \, \rmi \, r_{12} \omega_{12} \nonumber \\ + v_{12}^2 \biggl\{ \frac{5}{6} \biggl( 1 - \frac{19}{5} \nu \biggr) \dot{r}_{12} \pm \frac{1}{6} \bigl( 1 - 5 \nu \bigr) \rmi \, r_{12} \omega_{12} \biggr\} \biggr] e^{\mp \rmi \beta} \, , \\
		\fl \partial_t \widetilde{H}_1^{3,\pm 3} = \mp 4 \sqrt{\frac{5 \pi}{7}} \, \nu \, \frac{\delta M r_{12}}{r^3} \bigl( \dot{r}_{12} \mp \rmi \, r_{12} \omega_{12} \bigr)^2 \biggl[ 1 + \frac{5}{6} \biggl( 1 - \frac{19}{5} \nu \biggr) \dot{r}_{12}^2 \nonumber \\ \mp \frac{1}{3} \bigl( 1 - 2 \nu \bigr) \rmi \, r_{12} \omega_{12} \dot{r}_{12} + \frac{1}{2} \bigl( 1 - 5 \nu \bigr) r_{12}^2 \omega_{12}^2 \biggr) \biggr] e^{\mp 3 \rmi \beta} \, , \label{dtpolar2}
	\end{eqnarray}
\endnumparts
and for the odd perturbation
\numparts
	\begin{eqnarray}
		\fl \partial_t \widetilde{k}_0^{2,\pm 1} = \pm 2 \sqrt{\frac{2 \pi}{15}} \, \nu \, \frac{\delta M r_{12}}{r^2} \, r_{12} \omega_{12} \bigl( \dot{r}_{12} \mp \rmi \, r_{12} \omega_{12} \bigr) \biggl[ 1 + \frac{9}{14} \biggl( 1 - \frac{13}{3} \nu \biggr) v_{12}^2 \biggr] e^{\mp \rmi \beta} \, , \label{dtaxial1} \\
		\fl \partial_t \widetilde{k}_0^{3,0} = - 2 \sqrt{\frac{\pi}{7}} \, \nu \, \frac{M r_{12}^2}{r^3} \, r_{12} \omega_{12} \dot{r}_{12} \biggl[ \bigl( 1 - 3 \nu \bigr) + \biggl( \frac{5}{6} - \frac{37}{6} \nu + \frac{73}{6} \nu^2 \biggr) v_{12}^2 \biggr] \, , \\
		\fl \partial_t \widetilde{k}_0^{3,\pm 2} = \frac{2}{3} \sqrt{\frac{15 \pi}{14}} \, \nu \, \frac{M r_{12}^2}{r^3} r_{12} \omega_{12} \bigl( \dot{r}_{12} \mp \rmi \, r_{12} \omega_{12} \bigr) \biggl[ \bigl( 1 - 3 \nu \bigr) + \biggl( \frac{5}{6} - \frac{37}{6} \nu + \frac{73}{6} \nu^2 \biggr) \dot{r}_{12}^2 \nonumber \\ \mp \frac{4}{15} \bigl( 1 - 5 \nu + 5 \nu^2 \bigr) \rmi \, r_{12} \omega_{12} \dot{r}_{12} + \biggl( \frac{17}{30} - \frac{29}{6} \nu + \frac{65}{6} \nu^2 \biggr) r_{12}^2 \omega_{12}^2 \biggr] e^{\mp 2 \rmi \beta} \, , \\
		\fl \partial_t \widetilde{k}_1^{3,0} = \sqrt{\frac{\pi}{7}} \, \nu \, \frac{M r_{12}}{r^2} \bigl( 1 - 3 \nu \bigr) v_{12}^2 r_{12} \omega_{12} \, , \\
		\fl \partial_t \widetilde{k}_1^{3,\pm 2} = - \sqrt{\frac{5 \pi}{42}} \, \nu \, \frac{M r_{12}}{r^2} \bigl( 1 - 3 \nu \bigr) \bigl( \dot{r}_{12} \mp \rmi \, r_{12} \omega_{12} \bigr)^2  r_{12} \omega_{12} \, e^{\mp 2 \rmi \beta} \, . \label{dtaxial2}
	\end{eqnarray}
\endnumparts

We then check that the Einstein equations are satisfied for all $(\ell,m)$ up to $\ell = 3$ for a \textit{generic non circular} orbit, up to terms $\mathcal{O}(G^2,c^{-6},r_{12}^4)$. We give them here in the linear case for completeness (see e.g. \cite{Ze1.70,Gl.al.00,Sa.al.03} for general expressions). For the even perturbation, these seven equations read
\numparts
	\begin{eqnarray}
		\fl \partial_t \widetilde{H}_1^{\ell,m} - \partial_r \bigl( \widetilde{H}^{\ell,m} - \widetilde{K}^{\ell,m} \bigr) = 0 \, , \label{ein_pol1} \\
		\fl \partial_t \bigl( \widetilde{H}^{\ell,m} + \widetilde{K}^{\ell,m} \bigr) - \partial_r \widetilde{H}_1^{\ell,m} = 0 \, , \label{ein_pol2} \\
		\fl \partial_t \biggl[ \partial_r \widetilde{K}^{\ell,m} - \frac{1}{r} \bigl( \widetilde{H}^{\ell,m} - \widetilde{K}^{\ell,m} \bigr) \biggr] - \frac{\ell(\ell+1)}{2 r^2} \widetilde{H}_1^{\ell,m} = 0 \, , \\
		\fl \partial_t \biggl[ \partial_t \widetilde{K}^{\ell,m} - \frac{2}{r} \widetilde{H}_1^{\ell,m} \biggr] + \frac{1}{r} \partial_r \bigl( \widetilde{H}^{\ell,m} - \widetilde{K}^{\ell,m} \bigr) \nonumber \\ - \frac{(\ell-1)(\ell+2)}{2 r^2} \bigl( \widetilde{H}^{\ell,m} - \widetilde{K}^{\ell,m} \bigr) = 0 \, , \\
		\fl \partial^2_r \widetilde{K}^{\ell,m} - \frac{1}{r} \partial_r \bigl( \widetilde{H}^{\ell,m} - 3 \widetilde{K}^{\ell,m} \bigr) - \frac{(\ell-1)(\ell+2)}{2 r^2} \widetilde{K}^{\ell,m} \nonumber \\ - \biggl[ \frac{\ell(\ell+1)}{2} + 1 \biggr] \frac{\widetilde{H}^{\ell,m}}{r^2} = 0 \, , \\
		\fl \partial_t \biggl[ \partial_t \bigl( \widetilde{H}^{\ell,m} + \widetilde{K}^{\ell,m} \bigr) - 2 \biggl( \partial_r \widetilde{H}_1^{\ell,m} + \frac{\widetilde{H}_1^{\ell,m}}{r} \biggr) \biggr] + \partial^2_r \bigl( \widetilde{H}^{\ell,m} - \widetilde{K}^{\ell,m} \bigr) \nonumber \\ + \frac{2}{r} \partial_r \bigl( \widetilde{H}^{\ell,m} - \widetilde{K}^{\ell,m} \bigr) = 0 \, , \label{ein_pol6}
	\end{eqnarray}
\endnumparts
together with Eq.~\eref{H0H2}. Note that these equations are not all independent. For example, if Eqs.~\eref{ein_pol1} and \eref{ein_pol2} are satisfied, then Eq.~\eref{ein_pol6} is also satisfied. For the odd perturbation, the three remaining equations are
\numparts
	\begin{eqnarray}
		\fl \partial_t \widetilde{k}_0^{\ell,m} - \partial_r \widetilde{k}_1^{\ell,m} = 0 \, , \label{ein_ax1} \\
		\fl \partial_t \biggl[ \partial_r \widetilde{k}_1^{\ell,m} + \frac{2}{r} \widetilde{k}_1^{\ell,m} \biggr] - \partial^2_r \widetilde{k}_0^{\ell,m} + \frac{\ell(\ell+1)}{r^2} \widetilde{k}_0^{\ell,m} = 0 \, , \\
		\fl \partial_t \biggl[ \partial_t \widetilde{k}_1^{\ell,m} - \partial_r \widetilde{k}_0^{\ell,m} + \frac{2}{r} \widetilde{k}_0^{\ell,m} \biggr] + \frac{(\ell-1)(\ell+2)}{r^2} \widetilde{k}_1^{\ell,m} = 0 \, . \label{ein_ax3}
	\end{eqnarray}
\endnumparts
Assuming that all terms of a given equation in \eref{ein_pol1}--\eref{ein_pol6} and \eref{ein_ax1}--\eref{ein_ax3} are of the same order of magnitude, we can now understand by coming back to the initial metric decomposition \eref{polardecomp1}--\eref{polardecomp2} and \eref{axialdecomp1}--\eref{axialdecomp2} that it is necessary to expand $g^{\rm PN}_{00}$, $g^{\rm PN}_{0i}$ and $g^{\rm PN}_{ij}$ at the same PN order in Eqs.~\eref{metric2PN00}--\eref{metric2PNij}, i.e. up to $\mathcal{O}(c^{-5})$ included. The previous verification of the field equations provides a good check of the algebra yielding the perturbation coefficients \eref{polar1}--\eref{polar2} and \eref{axial1}--\eref{axial2}.

\section{Numerical evolution of the perturbation}
\label{secV}

\subsection{Regge-Wheeler and Zerilli master functions}\label{secVsubI}

From the multipolar coefficients $\widetilde{H}^{\ell,m}$, $\widetilde{H}_1^{\ell,m}$, $\widetilde{K}^{\ell,m}$, $\widetilde{k}_0^{\ell,m}$ and $\widetilde{k}_1^{\ell,m}$ one can construct for any $(\ell,m)$ two gauge-invariant scalar fields, namely the Regge-Wheeler \cite{ReWh.57} function $\Psi_{\ell,m}^{({\rm o})}$ and the Zerilli \cite{Ze1.70} function $\Psi_{\ell,m}^{({\rm e})}$, which contain all the information about the perturbation of the Schwarzschild metric. Gauge-invariant expressions of $\Psi_{\ell,m}^{({\rm e,o})}$ in terms of the multipolar coefficients in a general gauge are given e.g. in \cite{MaPo.05,NaRe.05,Ru.al.08}. In the Regge-Wheeler gauge \cite{ReWh.57}, the coefficients $\widetilde{G}^{\ell,m}$, $\widetilde{h}_0^{\ell,m}$, $\widetilde{h}_1^{\ell,m}$ and $\widetilde{k}_2^{\ell,m}$ vanish, so that these expressions get simplified and read
\numparts
	\begin{eqnarray}
		\Psi_{\ell,m}^{({\rm e})} & = \frac{r}{2(\lambda_\ell + 1)} \left( \widetilde{K}^{\ell,m} + \frac{r - 2M}{\lambda_\ell \, r + 3M} \left[ \widetilde{H}^{\ell,m} - r \partial_r \widetilde{K}^{\ell,m} \right] \right) , \label{Z_master} \\
		\Psi_{\ell,m}^{({\rm o})} & = \frac{r}{2 \lambda_\ell} \left( \partial_t \widetilde{k}_1^{\ell,m} - \partial_r \widetilde{k}_0^{\ell,m} + \frac{2}{r} \widetilde{k}_0^{\ell,m} \right) , \label{RW_master}
	\end{eqnarray}
\endnumparts
where we introduced the widely used notation $\lambda_\ell \equiv \frac{1}{2}(\ell-1)(\ell+2)$ \cite{Cha}. Note that the multipolar coefficent $\widetilde{H}_1^{\ell,m}$ does not enter the expression of $\Psi^{({\rm e})}_{\ell,m}$. Because we are considering linear perturbations, the master functions $\Psi_{\ell,m}^{({\rm e,o})}$ are defined up to a scale factor. We use the same convention as in \cite{NaRe.05}, emphasizing the link between $\Psi_{\ell,m}^{({\rm e,o})}$ and the polarization states $h_+$ and $h_\times$ of the gravitational waves at future null infinity; with our convention the two independent $+$ and $\times$ polarization states are given by
\begin{equation}\label{h_psi0}
	h_+ - \rmi \, h_\times = \frac{1}{r} \sum_{\ell,m} \sqrt{\frac{(\ell+2)!}{(\ell-2)!}} \left( \Psi_{\ell,m}^{({\rm e})} + \rmi \, \Psi_{\ell,m}^{({\rm o})} \right) {}_{-2}Y_{\ell,m} + \mathcal{O}(r^{-2}) \, ,
\end{equation}
where ${}_{-2}Y_{\ell,m}$ denotes the spin-weighted spherical harmonics of weight $-2$. The asymptotic waveform is also related to the more fundamental Weyl scalar $\Psi_4$, which admits a closed-form expression in terms of the master functions $\Psi^{({\rm e,o})}_{\ell,m}$ (see the Appendix).

The two master functions satisfy a wave equation with specific potentials $\mathcal{V}_\ell^{({\rm e,o})}$,
\begin{equation}\label{ZRW}
	\left( \partial^2_t - \partial^2_{r_*} + \mathcal{V}_\ell^{({\rm e,o})} \right) \Psi_{\ell,m}^{({\rm e,o})} = 0 \, ,
\end{equation}
where the so-called tortoise coordinate $r_*$ is related to the Schwarzschild radial coordinate $r$ by
\begin{equation}\label{r_star}
	r_* = r + 2M \, \textrm{ln} \left( \frac{r}{2M} - 1 \right) .
\end{equation}
Notice that these wave equations are only valid in vacuum; otherwise one has to include a source term in the right-hand-side of \eref{ZRW}, see e.g. \cite{MaPo.05,NaRe.05}. The Zerilli and Regge-Wheeler potentials read respectively
\begin{equation}
	\mathcal{V}_\ell^{({\rm e,o})} = \left( 1 - \frac{2M}{r} \right) \! \left( \frac{\ell	(\ell +1)}{r^2} - \frac{6M}{r^3} \, \mathcal{U}_\ell^{({\rm e,o})} \right) ,
\end{equation}
with
\numparts
	\begin{eqnarray}
			\mathcal{U}_\ell^{({\rm e})} &= \frac{\lambda_\ell (\lambda_\ell+2) r^2 + 3M (r-M)}{\left( \lambda_\ell \, r + 3M \right)^2} \, , \\ \mathcal{U}_\ell^{({\rm o})} &= 1 \, .
	\end{eqnarray}
\endnumparts
One can easily prove that $\frac{5}{7} < \mathcal{U}_\ell^{({\rm e})} < 2$ for all $\ell \geqslant 2$ and for all $r$ such that $2M < r < +\infty$, showing that the potentials $\mathcal{V}_\ell^{({\rm e})}$ and $\mathcal{V}_\ell^{({\rm o})}$ are very similar \cite{Cha}.

\subsection{Numerical evolution}\label{secVsubII}

The wave equations \eref{ZRW} are evolved numerically using the initial conditions at time $t=0$ and for any tortoise radius $r_*$, namely $\Psi_{\ell,m}^{({\rm e,o})}(0,r_*)$ and $\partial_t \Psi_{\ell,m}^{({\rm e,o})}(0,r_*)$, derived from the CL expansion of the 2PN metric for compact (i.e. point-mass) binaries. These initial conditions are calculated by plugging Eqs.~\eref{polar1}--\eref{polar2} and \eref{axial1}--\eref{axial2} and their partial time derivatives \eref{dtpolar1}--\eref{dtpolar2} and \eref{dtaxial1}--\eref{dtaxial2} into \eref{Z_master}--\eref{RW_master} and their partial time derivatives. 

We use Dirichlet boundary conditions, setting $\Psi_{\ell,m}^{({\rm e,o})}(t,r_*^{\rm min}) = \Psi_{\ell,m}^{({\rm e,o})}(t,r_*^{\rm max}) = 0$ at some radii $r_*^{\rm min}$ and $r_*^{\rm max}$. We choose the radii $r_*^{\rm min}$ and $r_*^{\rm max}$ in such a way that these boundary conditions are causaly disconnected from the computational domain, i.e. the spurious radiation generated on the boundaries $\{t,r_*^{\rm min,max}\}$ does not have time to propagate up to the extraction radius $r_*^{\rm ext} \equiv \frac{1}{2}(r_*^{\rm max} + r_*^{\rm min})$ for $0 \leqslant t \leqslant t^{\rm max}$, where $t^{\rm max} \equiv \frac{1}{2}(r_*^{\rm max} - r_*^{\rm min})$. Extending the computational domain to $[r_*^{\rm min},r_*^{\rm max}] \times \mathbb{R}_+$ would require using the Sommerfeld boundary conditions \cite{Som} $(\partial_t + \partial_{r_*}) \Psi_{\ell,m}^{({\rm e,o})}(t,r_*^{\rm max}) = 0$ and $(\partial_t - \partial_{r_*}) \Psi_{\ell,m}^{({\rm e,o})}(t,r_*^{\rm min}) = 0$, which are approximate boundary conditions, or even better some exact boundary conditions \cite{La1.04,La2.04}. The results below are based on computations where we have chosen $r_*^{\rm min} = -60 M$ and $r_*^{\rm max} = 660 M$, such that $r_*^{\rm ext} = 300 M$ and $t^{\rm max} = 360 M$.

A simple explicit second-order finite difference scheme has been used to evolve the wave equations \eref{ZRW}. We always choose the spatial grid resolution $\delta r_*$ and the time increment $\delta t$ such that the so-called Courant-Friedrichs-Lewy condition $\delta t < \delta r_*$ is verified; therefore the code is stable. The results in Figs~\ref{psi_po_head-on_BL}, \ref{psi_po_head-on}, \ref{psi_po} and \ref{psi_ax} below are based on computations where we used a spatial grid resolution $\delta r_* = 0.2 M$, and a time increment $\delta t = 0.1 M$.

\begin{figure}
	\begin{center}
		\includegraphics[width=10cm,angle=-90]{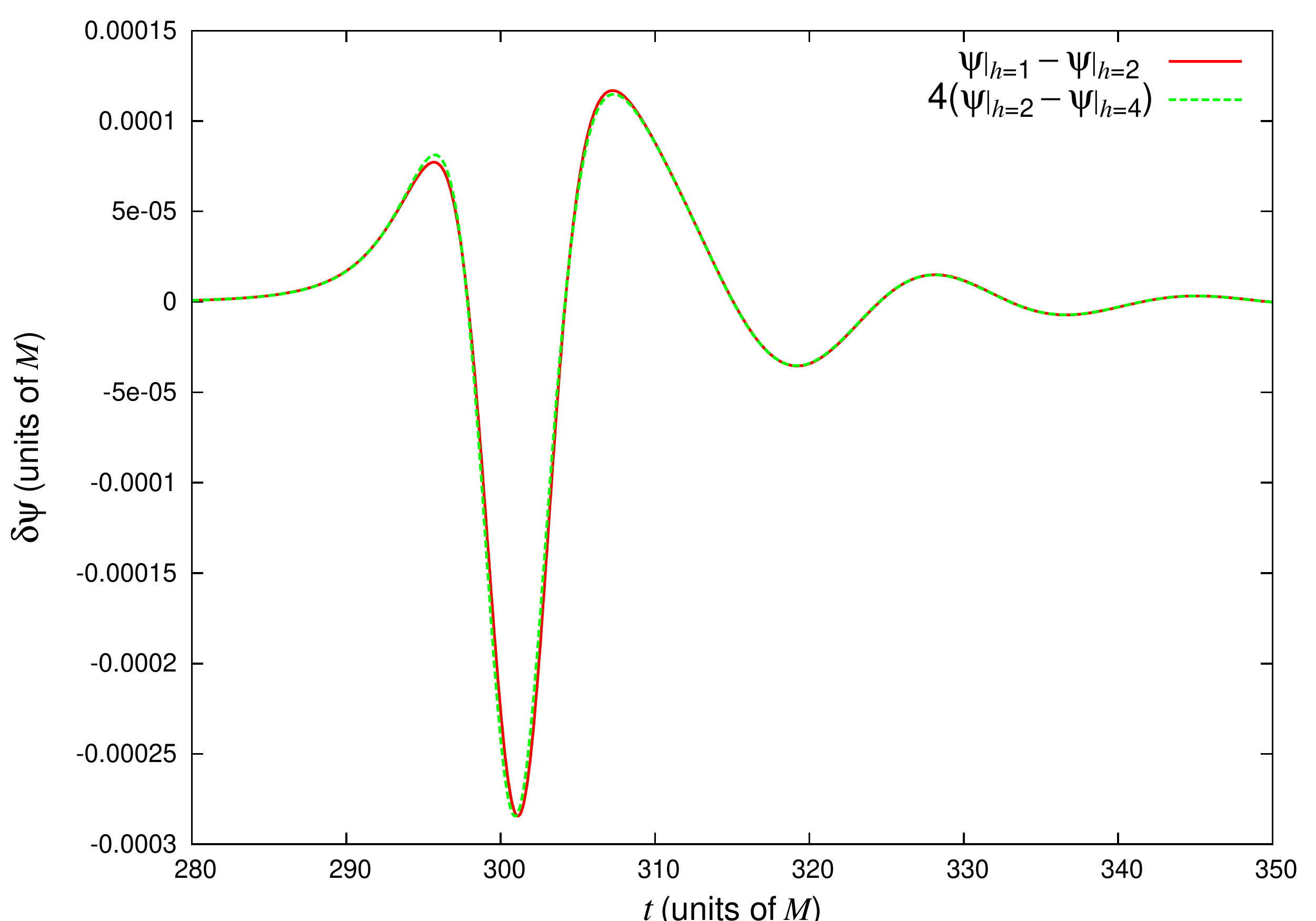}
		\caption{\footnotesize The good overlaping of the differences $\psi\vert_{h=1} - \psi\vert_{h=2}$ and $4(\psi\vert_{h=2} - \psi\vert_{h=4})$ demonstrates the second-order accuracy of the code.}
		\label{conv_res}
	\end{center}
\end{figure}

In order to check the second-order convergence of the code, we computed (for example) the real part of the $(\ell,m) = (2,2)$ mode of the Zerilli master function, $\psi \equiv \Re[\Psi^{({\rm e})}_{2,2}]$, for different spatial grid resolutions $\delta r_* = 0.2 M / h$, where $h=1,2,4$, with a constant time increment $\delta t = 0.025 M$, in the case of an unequal mass binary on circular orbit with $\nu = 0.185$, $r_{12} = 1.6 M$ and $\beta = 0$. The good overlapping of the differences $\psi\vert_{h=1} - \psi\vert_{h=2}$ and $4(\psi\vert_{h=2} - \psi\vert_{h=4})$ as shown in Fig.~\ref{conv_res} demonstrates the second-order accuracy of the code.

The numerical code was tested in several ways, and against previous published work as well as on some material presented in \cite{Le.al.09}:
\begin{enumerate}
	\item Using the Misner initial data \cite{Mi.60} as given by Price \& Pullin \cite{PrPu.94}, we reproduced in the case of head-on collision the waveform of their Fig.~2 and the associated radiated energy; cf. Eq.~(16) in \cite{PrPu.94}.
        \item Using the initial data provided by Sopuerta {\em et al.} \cite{So.al.06} (a conformally flat 3-metric with a Bowen-York extrinsic curvature and a Brill-Lindquist conformal factor), we reproduced their waveforms in Fig.~7, and the fluxes of energy, angular momentum and linear momentum of their Figs.~4, 5 and 8 respectively.\footnote{The updated plots of Fig.~7 and Fig.~8 of \cite{So.al.06} (taking into account the corrections from their first Erratum) are available on the e-Print server arXiv.org. Fig.~7 actually shows the Zerilli-Moncrief and Cunningham-Price-Moncrief master functions, and not their time derivatives as stated.}
        \item Checking that the total energy, angular momentum and linear momentum radiated do not depend on the physically irrelevant initial phase $\beta$; and that the components of the integrated linear momentum flux, or gravitational recoil, transform according to the usual law for vectors under a shift of the initial phase $\beta$.
        \item Checking that the quasi-normal mode frequencies of the waveforms are in good agreement with theoretical values \cite{No.99,KoSc.99,Be.al.09}.
\end{enumerate}

\subsection{Ringdown waveforms for head-on collisions}

\begin{figure}
	\begin{center}
		\includegraphics[width=10cm,angle=-90]{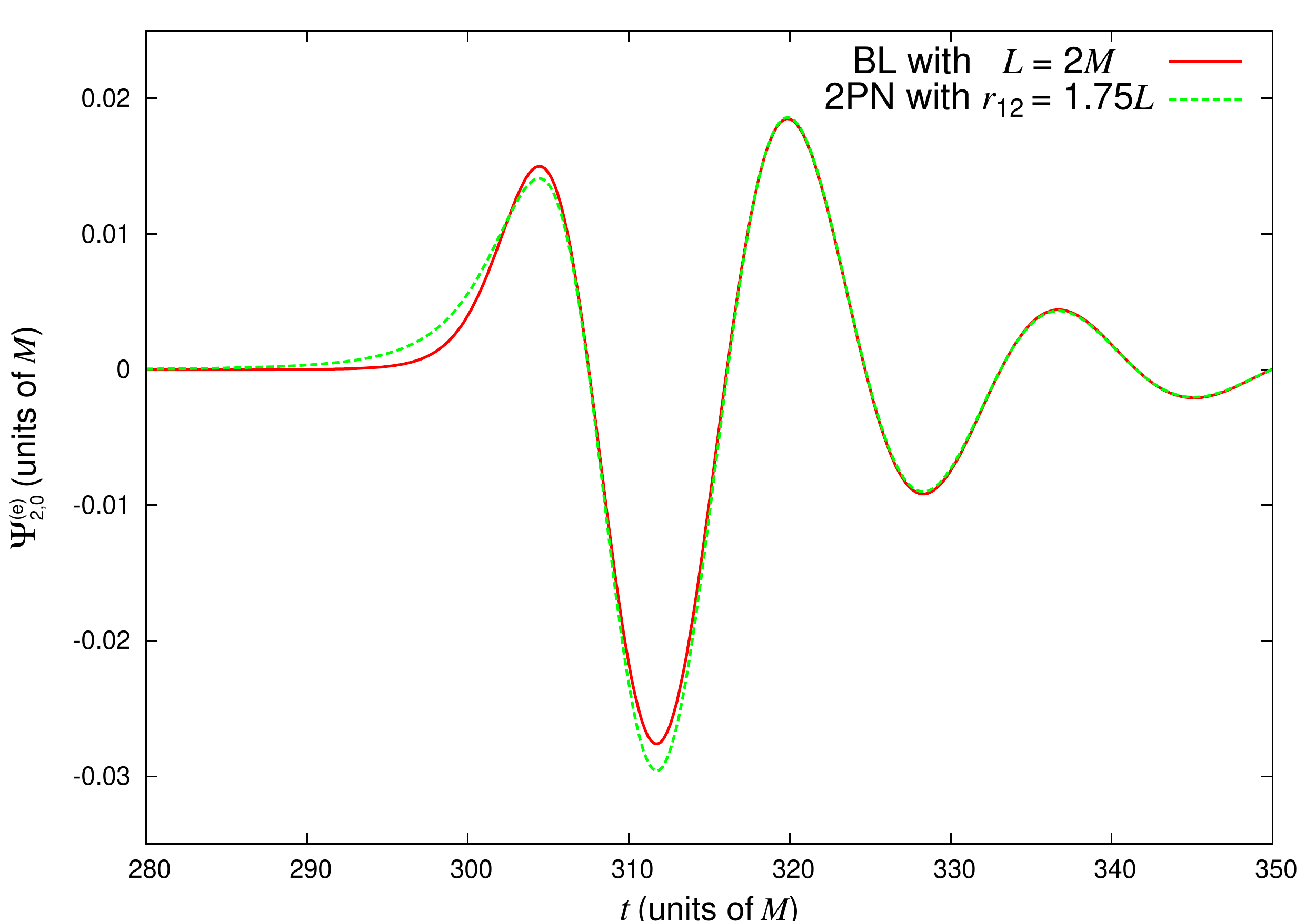}
		\caption{\footnotesize The real part of the mode $(2,0)$ of the Zerilli master function for even-parity (or polar) perturbations, in the case of an equal mass ($\nu = \frac{1}{4}$) head-on collision, using Brill-Lindquist initial data as given in Ref.~\cite{AbPr.96} with an initial distance $L = 2M$ (red), and our PN initial conditions with $r_{12} = 1.75 L$ (green).}
		\label{psi_po_head-on_BL}
	\end{center}
\end{figure}

\begin{figure}
	\begin{center}
		\includegraphics[width=10cm,angle=-90]{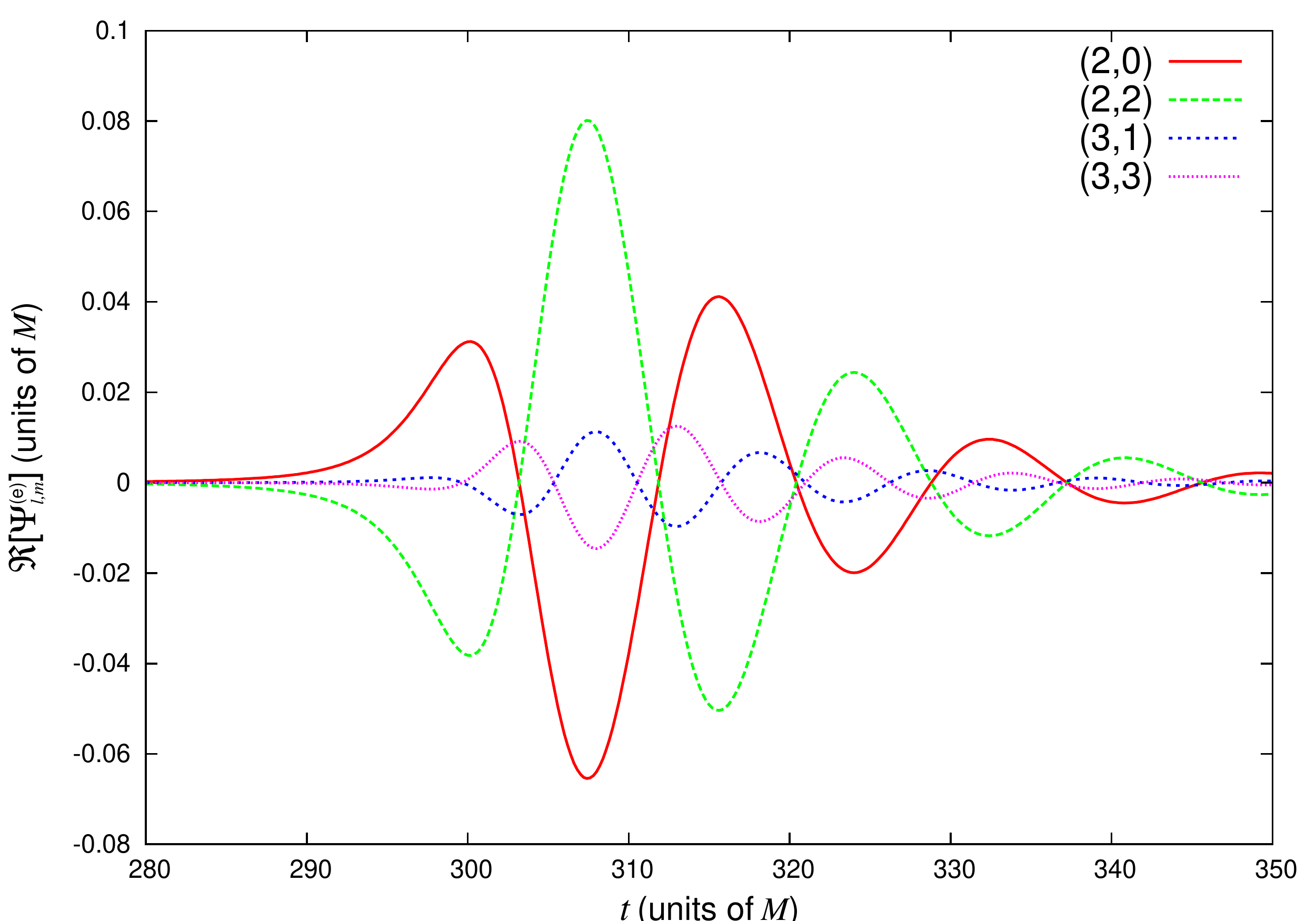}
		\caption{\footnotesize The real part of the non-vanishing modes $(\ell,m)$ of the Zerilli master function for even-parity (or polar) perturbations, up to $\ell = 3$, in the case of a head-on collision with $\nu = 0.2$ and $r_{12} = 4.4M$.}
		\label{psi_po_head-on}
	\end{center}
\end{figure}

We first consider head-on collisions for which the perturbation is purely polar, i.e. $\Psi_{\ell,m}^{\rm (o)} = 0$. We thus show in Figs.~\ref{psi_po_head-on_BL} and \ref{psi_po_head-on} the real part of the Zerilli master function for even-parity (or polar) perturbations, $\Re[\Psi^{({\rm e})}_{\ell,m}]$. Initial data for this case are obtained by setting $\omega_{12} = 0$ (no orbital angular momentum) and $\dot{r}_{12} = 0$ (time-symmetric initial conditions) in the expressions \eref{polar1}--\eref{polar2} and \eref{dtpolar1}--\eref{dtpolar2} of the even-parity multipolar coefficients and their partial time derivatives.

In Fig.~\ref{psi_po_head-on_BL} we consider an equal mass binary ($\nu = \frac{1}{4}$), and compare our 2PN-accurate results to those of Abrahams \& Price \cite{AbPr.96}, who studied similar head-on collisions using Brill-Lindquist (BL) initial data \cite{BrLi.63}. We shall restrict the comparison to the $(\ell,m)=(2,0)$ mode $\Psi^{({\rm e})}_{2,0}$ for simplicity. This comparison requires a detailed discussion of the relation between the two notions of distance between the two bodies used in both initial data sets. Using our conventions for the perturbation and various angles, we find that the only non-vanishing multipolar coefficients in the BL geometry are
\begin{equation}\label{H_BL}
	\widetilde{H}_{\rm BL}^{2,0} = \widetilde{K}_{\rm BL}^{2,0} = - \frac{1}{2} \sqrt{\frac{\pi}{5}} \frac{M L^2}{R^3} \frac{1}{1+\frac{M}{2R}} \, ,
\end{equation}
where $R = \frac{1}{4} \left( \sqrt{r} + \sqrt{r-2M} \right)^2$ is the isotropic radial coordinate, and $L$ is the distance between the two black holes of the BL solution. Because the multipolar coefficients $\widetilde{G}^{\ell,m}$, $\widetilde{h}_0^{\ell,m}$ and $\widetilde{h}_1^{\ell,m}$ vanish, the gauge in which the perturbation \eref{H_BL} is written coincides with the Regge-Wheeler gauge (hence our use of the symbol $\sim$ on the multipolar coefficients). Setting $\omega_{12} = 0$ and $\dot{r}_{12} = 0$ in \eref{polar1} and \eref{K20}, we get for our PN initial conditions
\begin{equation}\label{H_PN}
	\widetilde{H}_{\rm PN}^{2,0} = \widetilde{K}_{\rm PN}^{2,0} = - \frac{1}{2} \sqrt{\frac{\pi}{5}} \frac{M r_{12}^2}{r^3} = - \frac{1}{2} \sqrt{\frac{\pi}{5}} \frac{M r_{12}^2}{R^3} \frac{1}{\left(1+\frac{M}{2R}\right)^6} \, ,
\end{equation}
where we used $r = R \left( 1 + \frac{M}{2R} \right)^2$. Observe first that if we set $r_{12} = L$, then the two perturbations \eref{H_BL} and \eref{H_PN} coincide as they should in the weak-field domain $R \gg M$. But in the strong field domain $R \gtrsim M/2$, the comparison of the initial distances $L$ and $r_{12}$ is difficult. It then becomes interesting to check if these two measures of the distance between the holes can be related in such a way that the two waveforms compare well.

Most of the perturbation $\Psi^{({\rm e})}_{2,0}$ that propagates to future null infinity is generated around the maximum of the $\ell=2$ potential $\mathcal{V}_2^{\rm (e)}$ for polar perturbations, which is located around $r \simeq 3.1M$, or in terms of the isotropic coordinate $R \simeq 2M$. If we wish to identify the perturbations \eref{H_BL} and \eref{H_PN}, it is then natural to impose the \textit{definition}
\begin{equation}\label{def}
	\frac{r_{12}}{L} \equiv \left( 1 + \frac{M}{2R} \right)^{5/2} \Bigg|_{R \simeq 2M} \simeq 1.75 \, .
\end{equation}
We show in Fig.~\ref{psi_po_head-on_BL} the $(\ell,m)=(2,0)$ mode of the Zerilli master function using both BL initial data with $L = 2M$, and our PN initial conditions with $r_{12} = 1.75L$. We observe that the waveforms compare very well, which means that our post-Newtonian initial conditions are essentially equivalent to the Brill-Lindquist initial data in the case of head-on collisions. We find that the waveform computed with PN initial conditions is slightly delayed with respect to the BL one. So in order to achieve this agreement we also had to translate in time the PN curve by an amount $\Delta t \simeq 4M$. We checked that this good agreement does not depend on the value of the initial distance $L$.

We now focus on the waveforms obtained from our CL approximation with 2PN-accurate initial data. In Fig.~\ref{psi_po_head-on} we consider the head-on collision (with time-symmetric initial conditions) of an \textit{unequal} mass binary with mass ratio $\nu = 0.2$, and show the real part of the non-vanishing modes of the Zerilli master function for even-parity perturbations, $\Re[\Psi^{({\rm e})}_{\ell,m}]$, up to $\ell = 3$. We set the initial distance to a value twice larger than in the orbital case shown later, i.e. $r_{12} = 4.4M$, so that the Newtonian energy of the binary $E = - G m_1 m_2 / r_{12}$ is identical to that of the circular orbit configuration with initial distance $r_{12} = 2.2M$. This configuration being axisymmetric, for a given $\ell$ all $(\ell,m)$ modes can be related to the $(\ell,0)$ mode. For example we know from Ref.~\cite{Gu.al.08} that $\Psi^{({\rm e})}_{2,2} = -\frac{\sqrt{6}}{2} \Psi^{({\rm e})}_{2,0}$ and $\Psi^{({\rm e})}_{3,3} = -\sqrt{\frac{5}{3}} \Psi^{({\rm e})}_{3,1}$. We checked that the ratios of the amplitudes of the modes shown in Fig.~\ref{psi_po_head-on} are in very good agreement with these theoretical values.

\subsection{Ringdown waveforms for circular orbits}

In the case of circular orbits, both even and odd perturbations contribute. In this case we simply have to set $\dot{r}_{12} = 0$ in Eqs.~\eref{polar1}--\eref{polar2}, \eref{axial1}--\eref{axial2}, \eref{dtpolar1}--\eref{dtpolar2} and \eref{dtaxial1}--\eref{dtaxial2}. Our initial data will therefore depend only on the (physically irrelevant) initial orbital phase $\beta$, the initial distance $r_{12}$ and the initial orbital frequency $\omega_{12}$. We know from the 1PN-accurate equations of motion that, for a circular orbit, the orbital frequency is related to the binary's separation by the Kepler-like law \cite{Bl.06}
\begin{equation}\label{omega2}
	\omega_{12}^2 = \frac{M}{r_{12}^3} \left[ 1 + ( \nu - 3 ) \frac{M}{r_{12}} \right] + \mathcal{O}(c^{-4}) \, ,
\end{equation}
so we have only one free parameter $r_{12}$. This initial orbital distance $r_{12}$ will be an important parameter since it will be used in applications like \cite{Le.al.09} as a ``\textit{matching radius}'' to connect the computation of the ringdown phase to the previous inspiral and/or plunge phases.

An important point is worth emphasizing at this stage. Recall that in our previous calculation of the initial data for the Regge-Wheeler and Zerilli equations we have systematically and consistently neglected the non-linear terms $\mathcal{O}(G^2)$. Thus the perturbation coefficients \eref{polar1}--\eref{polar2} and \eref{axial1}--\eref{axial2} we consider, and which are valid for general orbits, are linear. Now, by introducing the expression \eref{omega2} of the orbital frequency $\omega_{12}$ (where $M \equiv G M)$ into the results \eref{polar1}--\eref{polar2}, \eref{axial1}--\eref{axial2}, \eref{dtpolar1}--\eref{dtpolar2} and \eref{dtaxial1}--\eref{dtaxial2} for the multipoles and their time derivatives, we do obtain terms which are of order $\mathcal{O}(G^2)$ or more in the case of circular orbits. Those terms have to be kept as they are, because the result \eref{omega2} comes from an independent calculation at the level of the equations of motion. That is, once we have proved (in Sec.~\ref{secIV}) that the Einstein field equations are satisfied for \textit{generic} non-circular orbits, we are allowed to reduce the solution to the particular case of a circular orbit by inserting \eref{omega2}; our point is that this adds new powers of $G$ which constitute an integral part of our solution of the field equations. This being said, a more involved calculation making use of the theory of second-order perturbations of a Schwarzschild black hole would introduce other terms of the same order $G^2$ in the final solution. But we do not have access to these terms in this work, which is based on first-order perturbations.

\begin{figure}
	\begin{center}
		\includegraphics[width=10cm,angle=-90]{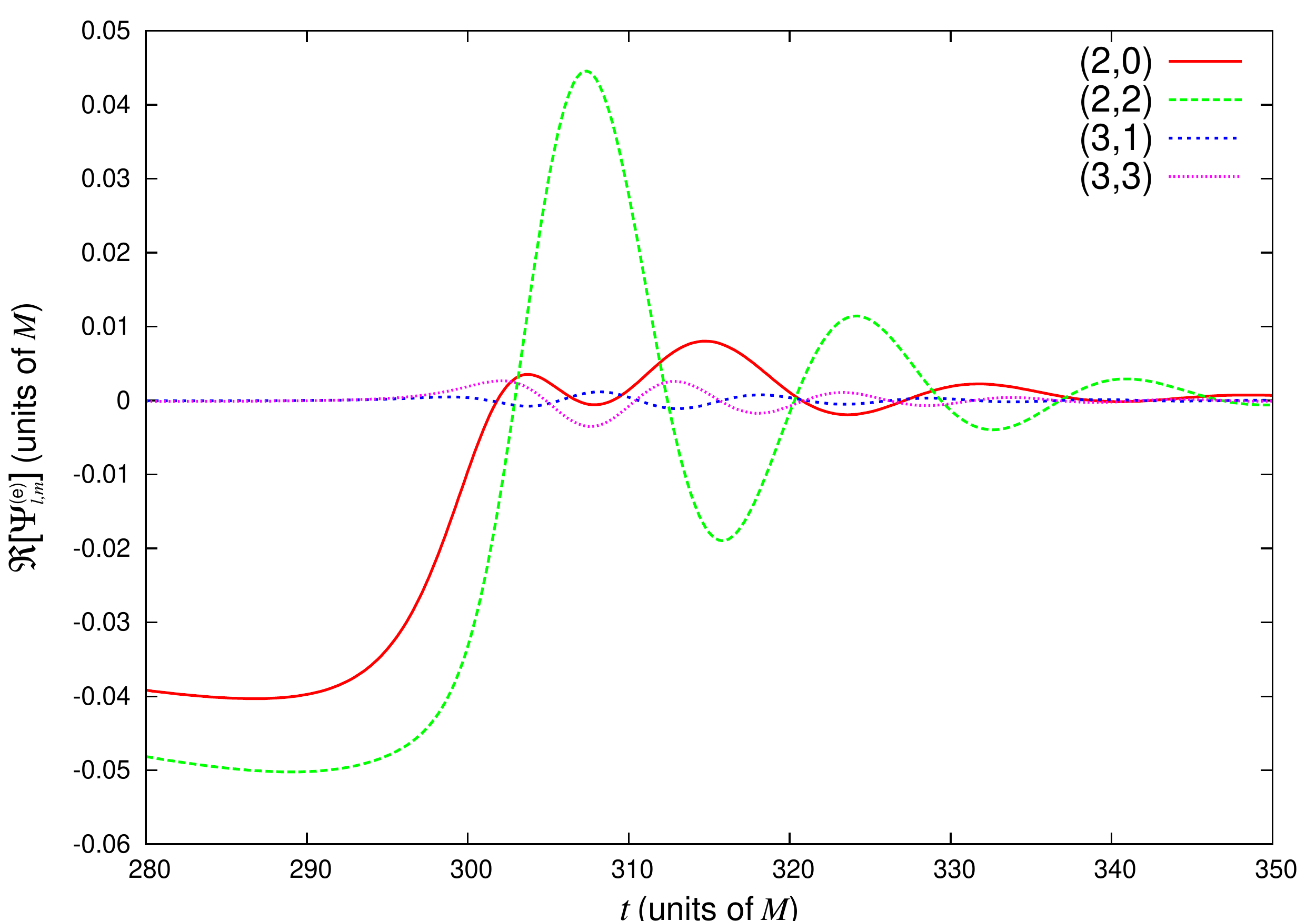}
		\caption{\footnotesize The real part of the non-vanishing modes $(\ell,m)$ of the Zerilli master function for even-parity (or polar) perturbations, up to $\ell = 3$, in the case of a circular orbit with $\nu = 0.2$, $r_{12} = 2.2M$ and $\beta = 0$.}
		\label{psi_po}
	\end{center}
\end{figure}
\begin{figure}
	\begin{center}
		\includegraphics[width=10cm,angle=-90]{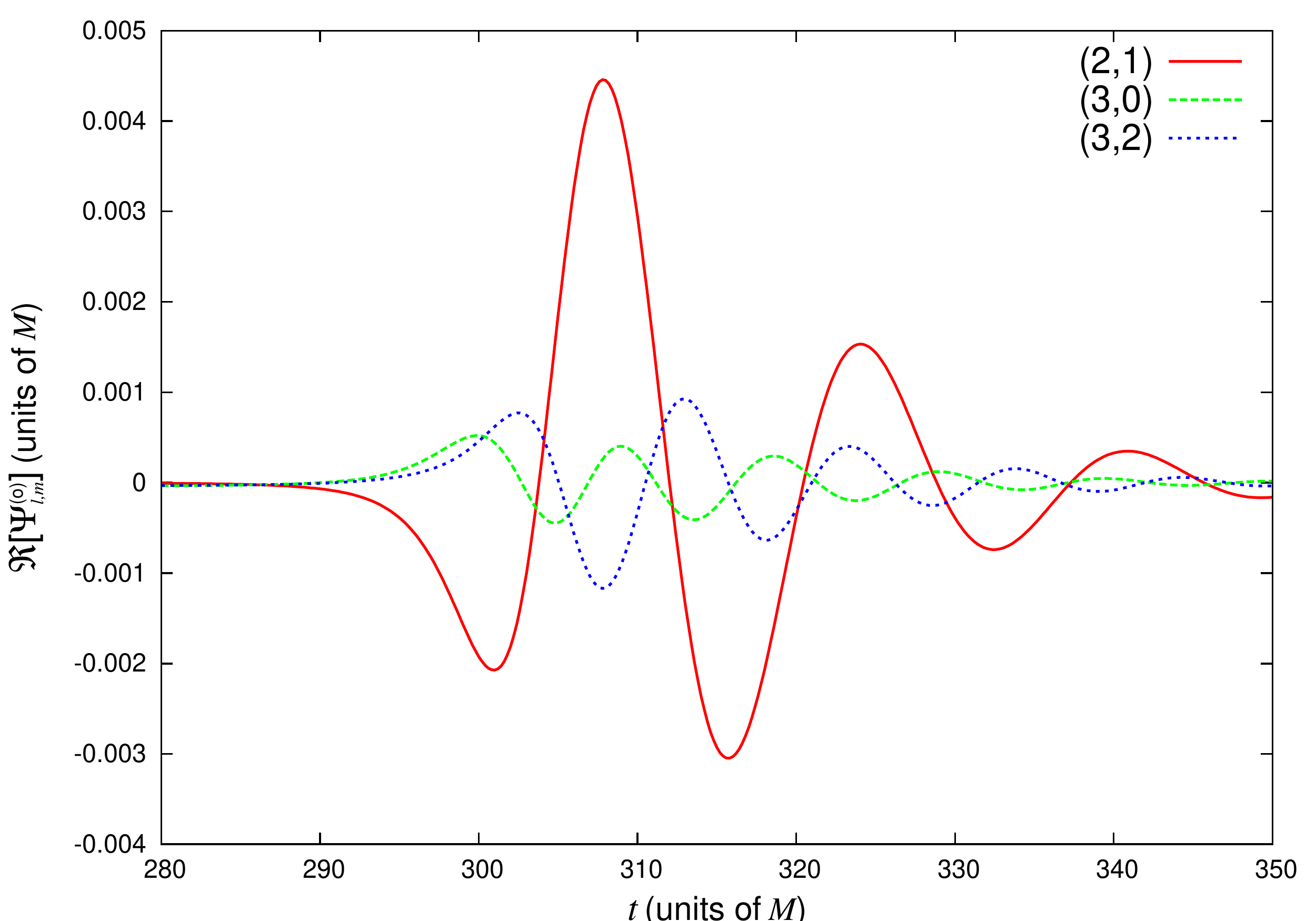}
		\caption{\footnotesize The real part of the non-vanishing modes $(\ell,m)$ of the Regge-Wheeler master function for odd-parity (or axial) perturbations, up to $\ell = 3$, in the case of a circular orbit with $\nu = 0.2$, $r_{12} = 2.2M$ and $\beta = 0$.}
		\label{psi_ax}
	\end{center}
\end{figure}

We show in Fig.~\ref{psi_po} the real part of the non-vanishing modes of the Zerilli master function for even-parity perturbations, $\Re[\Psi^{({\rm e})}_{\ell,m}]$, up to $\ell = 3$, in the case of the unequal mass binary with mass ratio $\nu = 0.2$ (same as before), on a circular orbit with initial separation $r_{12} = 2.2 M$ and initial phase $\beta = 0$. Notice the peculiar behaviour of the $\ell = 2$ modes which do not vanish asymptotically. This is because some terms in the CL-expanded 2PN metric \eref{metricPNCL00}--\eref{metricPNCLij} behave as $1/r$ in the far zone, therefore yielding non-vanishing asymptotic values for the quadrupolar modes of the Zerilli master function [see e.g. the third term in $h_{00}$ in Eq.~\eref{metricPNCL00}].

The Newtonian energy of the binary on a circular orbit in Fig.~\ref{psi_po} is chosen to be the same as for the head-on collision shown in Fig.~\ref{psi_po_head-on}. Notice the difference in amplitude, which is typically a factor 2, between the orbital case and the head-on case; the even (polar) perturbation is stronger in the head-on collision case than in the orbital case for the same total energy. This is consistent with the fact that in the case of a circular orbit, we have also in addition to the even-parity perturbations, some odd-parity or axial perturbations.

The odd/axial modes for the circular orbits are described by the Regge-Wheeler master function $\Psi^{({\rm o})}_{\ell,m}$. Notice that they can be computed with our PN initial conditions; Brill-Lindquist (BL) and Misner-Lindquist (ML) initial conditions are time-symmetric and can be applied only to the polar modes of head-on collisions. (See \cite{So.al.06,So.al.07} for initial conditions generalizing BL and ML, and which also permit to compute the axial modes of circular orbits in the CL approximation.) It is therefore particularly interesting to compute the real part of the Regge-Wheeler function, $\Re[\Psi^{({\rm o})}_{\ell,m}]$, as we do in Fig.~\ref{psi_ax} for the same unequal mass binary, and for the same initial conditions as for the polar modes shown in Fig.~\ref{psi_po}. Comparing Figs.~\ref{psi_po} and~\ref{psi_ax} we find that the amplitude of the even modes is approximately one order of magnitude larger than the amplitude of the odd modes. The even modes, which were the only ones present for head-on collisions without angular momentum, still dominate when we turn on the orbital angular momentum and consider the circular orbits.

\section{Conclusion}\label{secVI}

We have proposed an implementation of the close-limit (CL) approximation for binary black holes starting from post-Newtonian (PN) initial conditions developed at 2PN order. The 2PN metric generated by two point particles, when restricted to be linear in $G$ in order to be consistent \textit{in fine} with the linear black hole perturbation, was formally developed in CL form and identified with the metric of a linearly perturbed Schwarzschild black hole. We proved that the resulting multipolar coefficients describing the even-parity and odd-parity components of that perturbation consistently satisfy the perturbative Einstein field equations for general non-circular binary orbits. 

The post-Newtonian initial data was then specialized to the cases of head-on collisions and of circular orbits. We evolved numerically the Regge-Wheeler and Zerilli wave equations starting from those initial data, obtained the waveform generated during the ringdown phase, and compared these two cases. In a separate work \cite{Le.al.09}, we apply this formalism to the computation of the gravitational recoil produced during the ringdown phase of circular-orbit compact binaries, and match it to a previous PN calculation of the recoil accumulated in the preceding inspiral and merger phases \cite{Bl.al.05}.

They are several ways in which this work could be extended. First, one may expand to a higher order in the CL approximation, to get higher multipoles of the perturbation, or include in the initial 2PN metric terms describing the effect of spins of the initial black holes. We could also include terms $\mathcal{O}(G^2)$ or higher in the initial post-Newtonian metric, but it would be necessary to use the theory of at least \textit{second}-order perturbations of a Schwarzschild black hole \cite{Bri.al.09}. 

An important limitation of the present calculation is the impossibility to treat perturbations of a highly spinning black hole. In this work the initial orbital angular momentum of the binary had to be included in the perturbation of the final non-rotating black hole. A solution would be to employ the Teukolsky formalism \cite{Te.72} to perform similar calculations using a Kerr background instead of a Schwarzschild one. This would allow for a better description of the ringdown phase of the final black hole which is known from numerical calculations to be a rapidly spinning Kerr black hole \cite{Pr.05,Ca.al.06,Ba.al.06}. 

\ack

It is a pleasure to thank Bernard Whiting and Larry Price for their help in relating the Regge-Wheeler and Zerilli functions to the Weyl scalar $\Psi_4$. We thank also Emanuele Berti, Alessandra Buonanno, Luciano Rezzola and Clifford Will for very useful discussions. Algebraic computations were done with the software \textit{Mathematica}.

\appendix
\section{Basic material for black hole perturbations}\label{appA}

Although the material contained in this Appendix is well-known, we give in self-contained form the usual decomposition of an arbitrary linear perturbation $h_{\alpha\beta}$ of a Schwarzschild black hole onto the Zerilli-Mathews tensor spherical harmonics (correcting misprints frequently found in the litterature), and recall the relation between the Weyl scalar $\Psi_4$ and the Regge-Wheeler and Zerilli master functions, from which we derive the asymptotic waveform in the form of its two polarization states $+$ and $\times$.

\subsection{The basis of tensorial spherical harmonics}

In order to perform practical calculations, it is particularly convenient, instead of using the decomposition \eref{polardecomp1}--\eref{polardecomp2} and \eref{axialdecomp1}--\eref{axialdecomp2} introduced in Sec.~\ref{secIII}, to introduce the Zerilli-Mathews basis of tensor spherical harmonics $( e_{\alpha\beta}^{A,\ell,m} )_{A=1,\cdots,10}$, and to write the perturbation as
\begin{equation}
	h_{\alpha\beta}(t,r,\theta,\varphi) = \sum_{A=1}^{10} \sum_{\ell,m} h_{A,\ell,m}(t,r) \, e_{\alpha\beta}^{A,\ell,m}(\theta,\varphi) \, ,
\end{equation}
where $x^\alpha=\{t,r,\theta,\varphi\}$ are the usual Schwarzschild spherical coordinates. This is because the tensor spherical harmonics $( e_{\alpha\beta}^{A,\ell,m} )_{A=1,\cdots,10}$ form an orthonormal basis \cite{Ma.62,Ze1.70,Ze2.70}, in the sense that
\begin{equation}
	\langle e^{A,\ell,m} \cdot e^{A',\ell',m'} \rangle \equiv \int \xi^{\alpha\beta} \xi^{\gamma\delta} \, \bar{e}^{A,\ell,m}_{\alpha \gamma} \,
e^{A',\ell',m'}_{\beta \delta} \, \rmd \Omega = \delta_{AA'} \, \delta_{\ell\ell'} \, \delta_{mm'} \, ,
\end{equation}
where $\xi_{\alpha\beta}$ is the \textit{Euclidean} metric written in spherical coordinates, $\delta_{ij}$ is the usual Kronecker symbol, the overbar denotes complex conjugation, and the integration is performed on the sphere of unit radius. This basis being orthonormal, each component $h_{A,\ell,m}$ of a given perturbation $h_{\alpha\beta}$ can be calculated by the projection
\begin{equation}\label{scal_prod}
	h_{A,\ell,m} = \langle e^{A,\ell,m} \cdot h \rangle \, ,
\end{equation}
from which we then deduce the multipolar coefficients $H_0^{\ell,m},H_1^{\ell,m},\cdots,k_2^{\ell,m}$ defined by \eref{h1_H}--\eref{h10_H} below. This method was systematically used to get the results \eref{polar1}--\eref{polar2} and \eref{axial1}--\eref{axial2} from the information on the metric perturbation contained in Eqs.~\eref{metricPNCL00}--\eref{metricPNCLij}.

The Zerilli-Mathews basis of tensorial harmonics explicitly reads
\begin{eqnarray}
	e^{1,\ell,m}_{\alpha\beta} & = \left(
	\begin{array}{cccc}
		1 & 0 & 0 & 0 \\
		0 & 0 & 0 & 0 \\
		0 & 0 & 0 & 0 \\
		0 & 0 & 0 & 0
	\end{array} \right) Y_{\ell,m} \, , \\
		e^{2,\ell,m}_{\alpha\beta} & = \frac{\rmi}{\sqrt{2}} \left(
	\begin{array}{cccc}
		0 & 1 & 0 & 0 \\
		1 & 0 & 0 & 0 \\
		0 & 0 & 0 & 0 \\
		0 & 0 & 0 & 0
	\end{array} \right) Y_{\ell,m} \, , \\
	e^{3,\ell,m}_{\alpha\beta} & = \left(
	\begin{array}{cccc}
		0 & 0 & 0 & 0 \\
		0 & 1 & 0 & 0 \\
		0 & 0 & 0 & 0 \\
		0 & 0 & 0 & 0
	\end{array} \right) Y_{\ell,m} \, , \\
	e^{4,\ell,m}_{\alpha\beta} & = \rmi \sqrt{\frac{1}{2} \frac{(\ell-1)!}{(\ell+1)!}} \left(
	\begin{array}{cccc}
		0 & 0 & \partial_\theta & \partial_\varphi \\
		0 & 0 & 0 & 0 \\
		\partial_\theta & 0 & 0 & 0 \\
		\partial_\varphi & 0 & 0 & 0
	\end{array} \right) Y_{\ell,m} \, , \\
	e^{5,\ell,m}_{\alpha\beta} & = \sqrt{\frac{1}{2} \frac{(\ell-1)!}{(\ell+1)!}} \left(
	\begin{array}{cccc}
		0 & 0 & 0 & 0 \\
		0 & 0 & \partial_\theta & \partial_\varphi \\
		0 & \partial_\theta & 0 & 0 \\
		0 & \partial_\varphi & 0 & 0
	\end{array} \right) Y_{\ell,m} \, , \\
	e^{6,\ell,m}_{\alpha\beta} & = \frac{1}{\sqrt{2}} \left(
	\begin{array}{cccc}
		0 & 0 & 0 & 0 \\
		0 & 0 & 0 & 0 \\
		0 & 0 & 1 & 0 \\
		0 & 0 & 0 & s^2
	\end{array} \right) Y_{\ell,m} \, , \\
	e^{7,\ell,m}_{\alpha\beta} & = \sqrt{\frac{1}{2} \frac{(\ell-2)!}{(\ell+2)!}} \left(
	\begin{array}{cccc}
		0 & 0 & 0 & 0 \\
		0 & 0 & 0 & 0 \\
		0 & 0 & \mathcal{D}_2 & \mathcal{D}_1 \\
		0 & 0 & \mathcal{D}_1 & -s^2 \, \mathcal{D}_2
	\end{array} \right) Y_{\ell,m} \, , \\
	e^{8,\ell,m}_{\alpha\beta} & = \sqrt{\frac{1}{2}	\frac{(\ell-1)!}{(\ell+1)!}} \left(
	\begin{array}{cccc}
		0 & 0 & s^{-1} \, \partial_\varphi & -s \, \partial_\theta \\
		0 & 0 & 0 & 0 \\
		s^{-1} \, \partial_\varphi & 0 & 0 & 0 \\
		-s \, \partial_\theta & 0 & 0 & 0
	\end{array} \right) Y_{\ell,m} \, , \\
	e^{9,\ell,m}_{\alpha\beta} & = \rmi \sqrt{\frac{1}{2}	\frac{(\ell-1)!}{(\ell+1)!}} \left(
	\begin{array}{cccc}
		0 & 0 & 0 & 0 \\
		0 & 0 & s^{-1} \, \partial_\varphi & -s \, \partial_\theta \\
		0 & s^{-1} \, \partial_\varphi & 0 & 0 \\
		0 & -s \, \partial_\theta & 0 & 0
	\end{array} \right) Y_{\ell,m} \, , \\
	e^{10,\ell,m}_{\alpha\beta} & = \rmi \sqrt{\frac{1}{2} \frac{(\ell-2)!}{(\ell+2)!}} \left(
	\begin{array}{cccc}
		0 & 0 & 0 & 0 \\
		0 & 0 & 0 & 0 \\
		0 & 0 &	s^{-1} \mathcal{D}_1 & -s \, \mathcal{D}_2 \\
		0 & 0 & -s \, \mathcal{D}_2 & -s \, \mathcal{D}_1
	\end{array} \right) Y_{\ell,m} \, ,
\end{eqnarray}
where we introduced the convenient shortcut $s \equiv \sin{\theta}$, together with the angular operators
\begin{eqnarray}
	\mathcal{D}_1 & \equiv 2 \left( \partial_\theta - \cot{\theta} \right) \partial_\varphi \, , \\
	\mathcal{D}_2 & \equiv \partial^2_\theta - \cot{\theta} \, \partial_\theta - s^{-2} \, \partial^2_\varphi \, ,
\end{eqnarray}
and where our convention regarding the spherical harmonics is
\begin{equation}\label{Ylm}
Y_{\ell,m} (\theta,\varphi) \equiv \sqrt{\frac{2 \ell + 1}{4 \pi}\frac{(\ell-m)!}{(\ell+m)!}} \, P_{\ell,m}(\cos{\theta}) \, e^{\rmi m \varphi} \, ,
\end{equation}
with $P_{\ell,m}(x)=(-1)^m (1-x^2)^{m/2} \bigl(\frac{\rmd}{\rmd x}\bigr)^m P_\ell(x)$ being the associated Legendre functions.

Finally, the coefficients $h_{A,\ell,m}$ of the arbitrary perturbation $h_{\alpha \beta}$ are related to the multipolar coefficients $H_0^{\ell,m},H_1^{\ell,m},\cdots,k_2^{\ell,m}$ of the decomposition \eref{polardecomp1}--\eref{polardecomp2} and \eref{axialdecomp1}--\eref{axialdecomp2} through
\begin{eqnarray}
	h_{1,\ell,m} & = \left( 1 - \frac{2M}{r} \right) H_0^{\ell,m} \, , \label{h1_H} \\
	h_{2,\ell,m} & = - \rmi \sqrt{2} \, H_1^{\ell,m} \, , \\
	h_{3,\ell,m} & = \left( 1 - \frac{2M}{r} \right)^{-1} H_2^{\ell,m} \, , \\
	h_{4,\ell,m} & = - \rmi \sqrt{2 \frac{(\ell+1)!}{(\ell-1)!}} \, h_0^{\ell,m} \, , \\
	h_{5,\ell,m} & = \sqrt{2 \frac{(\ell+1)!}{(\ell-1)!}} \, h_1^{\ell,m} \, , \\
	h_{6,\ell,m} & = r^2 \sqrt{2} \left( K^{\ell,m} - \frac{\ell(\ell+1)}{2} G^{\ell,m} \right) , \\
	h_{7,\ell,m} & = r^2 \sqrt{\frac{1}{2} \frac{(\ell+2)!}{(\ell-2)!}} \, G^{\ell,m} \, , \\
	h_{8,\ell,m} & = - \sqrt{2 \frac{(\ell+1)!}{(\ell-1)!}} \, k_0^{\ell,m} \, , \\
	h_{9,\ell,m} & = \rmi \sqrt{2 \frac{(\ell+1)!}{(\ell-1)!}} \, k_1^{\ell,m} \, , \\
	h_{10,\ell,m} & = - \rmi \sqrt{\frac{1}{2} \frac{(\ell+2)!}{(\ell-2)!}} \, k_2^{\ell,m} \label{h10_H} \, .
\end{eqnarray}

\subsection{Link between the asymptotic waveform and the master functions}
\label{appB}

Here we recall the general expression (known as a Chandrasekhar transformation \cite{Cha}) of the Weyl scalar $\Psi_4$ in terms of the Regge-Wheeler and Zerilli master functions $\Psi_{\ell,m}^{({\rm e,o})}$, and we compute the combination $h_+ - \rmi \, h_\times$, where $h_{+,\times}$ denote the two asymptotic wave polarizations.

In the Schwarzschild spherical coordinate system $\{ t,r,\theta,\varphi \}$, we let $(e^\alpha_r,e^\alpha_\theta,e^\alpha_\varphi)$ be the associated orthonormal basis of $t = \textrm{const}$ hypersurfaces, and $e^\alpha_t$ be the timelike unit vector orthogonal to them. Then, we consider the following complex null tetrad: $k^\alpha$ and $l^\alpha$ are two real null vectors, while $m^\alpha$ and $\bar{m}^\alpha$ are complex conjugated null vectors defined by
\begin{eqnarray}
	k^\alpha &= \frac{1}{\sqrt{2}} \left( e^\alpha_t + e^\alpha_r \right) , \\
	l^\alpha &= \frac{1}{\sqrt{2}} \left( e^\alpha_t - e^\alpha_r \right) , \\
	m^\alpha &= \frac{1}{\sqrt{2}} \left( e^\alpha_\theta - \rmi \, e^\alpha_\varphi \right) , \\
	\bar{m}^\alpha &= \frac{1}{\sqrt{2}} \left( e^\alpha_\theta + \rmi \,e^\alpha_\varphi \right) .
\end{eqnarray}
We therefore have $m^\alpha \bar{m}_\alpha = 1 = - l^\alpha k_\alpha$, all the other scalar products vanishing. Various conventions for the definition of the Weyl scalar $\Psi_4$ can be found in the litterature. We adopt
\begin{equation}\label{def_Psi4}
	\Psi_4 \equiv C_{\alpha \beta \gamma \delta} \, l^\alpha \,\bar{m}^\beta \,l^\gamma \,\bar{m}^\delta \, ,
\end{equation}
where $C_{\alpha \beta \gamma \delta}$ is the Weyl tensor, which coincides with the Riemann tensor in vacuum. It can be shown that, for a generic perturbation of a Schwarzschild black hole,
\begin{equation}\label{chandra}
	\Psi_4 = \frac{1}{r} \sum_{\ell,m} \sqrt{\frac{(\ell+2)!}{(\ell-2)!}} \left( \mathcal{P}_\ell^{({\rm e})} \,\Psi_{\ell,m}^{({\rm e})} + \rmi \,
\mathcal{P}_\ell^{({\rm o})} \,\Psi_{\ell,m}^{({\rm o})} \right) {}_{-2}Y_{\ell,m} \, ,
\end{equation}
where $r$ is the usual Schwarzschild radial coordinate, and the master functions $\Psi_{\ell,m}^{({\rm e,o})}$ are those defined in Eqs.~\eref{Z_master}--\eref{RW_master}. The relation \eref{chandra} is exact for first-order perturbations of the Schwarzschild geometry \cite{Cha}. The $s=-2$ spin-weighted spherical harmonics ${}_{-2}Y_{\ell,m}$ are defined for any integer $s$ by \cite{NePe.66,Go.al.67}
\begin{equation}\label{sYlm}
{}_{-s}Y_{\ell,m} (\theta,\varphi) \equiv (-1)^s \sqrt{\frac{2 \ell + 1}{4 \pi}} \, {}_{s}d_{\ell,m} (\theta) \, e^{\rmi m \varphi} \, ,
\end{equation}
where the Wigner functions ${}_{s}d_{\ell,m}$ read
\begin{equation}\label{sdlm}
	\fl {}_{s}d_{\ell,m} (\theta) \equiv \sum_{k = k_{\rm min}}^{k_{\rm max}} \frac{(-1)^k \sqrt{(\ell+m)! (\ell-m)! (\ell+s)! (\ell-s)!}}{k! (\ell+m-k)! (\ell-s-k)! (s-m+k)!} \left( \cos{\frac{\theta}{2}} \right)^{2\ell} \left( \tan{\frac{\theta}{2}} \right)^{2k+s-m} \hspace{-0.5cm} ,
\end{equation}
with $k_{\rm min} = {\rm max}(0,m-s)$ and $k_{\rm max} = {\rm min}(\ell+m,\ell-s)$. In the case $s=0$ we recover the scalar spherical harmonics \eref{Ylm}. The differential operators acting on $\Psi_{\ell,m}^{({\rm e,o})}$ in the relation \eref{chandra} read explicitly as
\begin{equation}
	\mathcal{P}_\ell^{({\rm e,o})} = \frac{1}{4} \Bigl( \mathcal{W}_\ell^{({\rm e,o})} + \partial_{r_*} - \partial_t \Bigr) \Bigl( \partial_{r_*} - \partial_t \Bigr) \, ,
\end{equation}
where $r_*$ is the tortoise coordinate \eref{r_star}, and the potentials $\mathcal{W}_\ell^{({\rm e,o})}$ are given by 
\begin{eqnarray}
	\mathcal{W}_\ell^{({\rm e})} & = \frac{2}{r} \left( 1 - \frac{3M}{r} \right) - \frac{6M(r-2M)}{r^2(\lambda_\ell \, r + 3M)} \, , \\
	\mathcal{W}_\ell^{({\rm o})} & = \frac{2}{r} \left( 1 - \frac{3M}{r} \right) \, .
\end{eqnarray}
Recall that $\lambda_\ell = \frac{1}{2}(\ell-1)(\ell+2)$.

Now, in the limit $r \rightarrow +\infty$, the Regge-Wheeler and Zerilli functions $\Psi_{\ell,m}^{({\rm e,o})}$ are functions of the retarded time $t-r_*$ only. Indeed, they are solutions of the wave equations \eref{ZRW} with decaying potentials $\mathcal{V}_\ell^{({\rm e,o})} \propto 1/r^2$. Because the potentials $\mathcal{W}_\ell^{({\rm e,o})}$ also decay as $r^{-1}$, we have the asymptotic expressions
\begin{equation}\label{asymp_exp}
	\mathcal{P}_\ell^{({\rm e,o})} \Psi_{\ell,m}^{({\rm e,o})} = \partial^2_t \Psi_{\ell,m}^{({\rm e,o})} + \mathcal{O}(r^{-1}) \, .
\end{equation}
Furthermore, working in the transverse and traceless gauge, and performing some projections onto the linearized Weyl tensor around flat spacetime in the definition \eref{def_Psi4} leads to the well-known expression
\begin{equation}\label{Psi4_h}
	\Psi_4 = \partial^2_t \bigl( h_+ - \rmi \, h_\times \bigr) + \mathcal{O}(r^{-2}) \, ,
\end{equation}
where the two polarization states $h_+$ and $h_\times$ are defined by
\begin{eqnarray}
	h_+ & \equiv \frac{1}{2} \left( e_\theta^i e_\theta^j - e_\varphi^i e_\varphi^j \right) h_{ij} = \frac{1}{2} \left( h_{\theta \theta} - h_{\varphi \varphi} \right) , \\
	h_\times & \equiv \frac{1}{2} \left( e_\theta^i e_\varphi^j + e_\varphi^i e_\theta^j \right) h_{ij} = h_{\theta \varphi} \, .
\end{eqnarray}
Finally, combining the results \eref{chandra} and \eref{Psi4_h} with the asymptotic expansion \eref{asymp_exp}, we recover the well-known formula
\begin{equation}\label{h_psi}
	h_+ - \rmi \, h_\times = \frac{1}{r} \sum_{\ell,m} \sqrt{\frac{(\ell+2)!}{(\ell-2)!}} \left( \Psi_{\ell,m}^{({\rm e})} + \rmi \, \Psi_{\ell,m}^{({\rm o})} \right) {}_{-2}Y_{\ell,m} + \mathcal{O}(r^{-2}) \, .
\end{equation}

\section*{References}

\bibliographystyle{iopart-num}
\bibliography{/tmp_mnt/netpapeur/users_home4/letiec/Publications/ListeRef}

\providecommand{\newblock}{}
\begin{thebibliography}{10}
\expandafter\ifx\csname url\endcsname\relax
  \def\url#1{{\tt #1}}\fi
\expandafter\ifx\csname urlprefix\endcsname\relax\def\urlprefix{URL }\fi
\providecommand{\eprint}[2][]{\url{#2}}

\bibitem{Bl.al.02}
Blanchet L, Faye G, Iyer B~R and Joguet B 2002 {\em Phys. Rev. D\/} {\bf 65}
  061501 (\textit{Preprint} \eprint{arXiv:gr-qc/0105099})

\bibitem{Bl2.al.04}
Blanchet L, Damour T, Esposito-Far\`ese G and Iyer B~R 2004 {\em Phys. Rev.
  Lett.\/} {\bf 93} 091101 (\textit{Preprint} \eprint{arXiv:gr-qc/0406012})

\bibitem{Bl.al.96}
Blanchet L, Iyer B~R, Will C~M and Wiseman A~G 1996 {\em Class. Quant. Grav.\/}
  {\bf 13} 575 (\textit{Preprint} \eprint{arXiv:gr-qc/9602024})

\bibitem{Ar.al.04}
Arun K~G, Blanchet L, Iyer B~R and Qusailah M~S~S 2004 {\em Class. Quant.
  Grav.\/} {\bf 21} 3771 (\textit{Preprint} \eprint{arXiv:gr-qc/0404085})

\bibitem{Bl.al.08}
Blanchet L, Faye G, Iyer B~R and Sinha S 2008 {\em Class. Quant. Grav.\/} {\bf
  25} 165003 (\textit{Preprint} \eprint{arXiv:0802.1249 [gr-qc]})

\bibitem{Bl.06}
Blanchet L 2006 {\em Living Rev. Rel.\/} {\bf 9} 4 (\textit{Preprint}
  \eprint{arXiv:gr-qc/0202016})

\bibitem{Pr.05}
Pretorius F 2005 {\em Phys. Rev. Lett.\/} {\bf 95} 121101 (\textit{Preprint}
  \eprint{arXiv:gr-qc/0507014})

\bibitem{Ca.al.06}
Campanelli M, Lousto C~O, Marronetti P and Zlochower Y 2006 {\em Phys. Rev.
  Lett.\/} {\bf 96} 111101 (\textit{Preprint} \eprint{arXiv:gr-qc/0511048})

\bibitem{Ba.al.06}
Baker J~G, Centrella J, Choi D~I, Koppitz M and van Meter J 2006 {\em Phys.
  Rev. Lett.\/} {\bf 96} 111102 (\textit{Preprint}
  \eprint{arXiv:gr-qc/0511103})

\bibitem{Bu.al.07}
Buonanno A, Cook G~B and Pretorius F 2007 {\em Phys. Rev. D\/} {\bf 75} 124018
  (\textit{Preprint} \eprint{arXiv:gr-qc/0610122})

\bibitem{Ba.al.07}
Baker J~G, van Meter J~R, McWilliams S~T, Centrella J and Kelly B~J 2007 {\em
  Phys. Rev. Lett.\/} {\bf 99} 181101 (\textit{Preprint}
  \eprint{arXiv:gr-qc/0612024})

\bibitem{Be.al.07}
Berti E, Cardoso V, Gonzalez J~A, Sperhake U, Hannam M, Husa S and Bru\"gmann B
  2007 {\em Phys. Rev. D\/} {\bf 76} 064034 (\textit{Preprint}
  \eprint{arXiv:gr-qc/0703053})

\bibitem{Pa.al.07}
Pan Y, Buonanno A, Baker J~G, Centrella J, Kelly B~J, McWilliams S~T, Pretorius
  F and van Meter J~R 2007 {\em Phys. Rev. D\/} {\bf 77} 024014
  (\textit{Preprint} \eprint{arXiv:0704.1964 [gr-qc]})

\bibitem{Ha.al.07}
Hannam M, Husa S, Gonz\'alez J~A, Sperhake U and Br\"ugmann B 2007 {\em Phys.
  Rev. D\/} {\bf 77} 044020 (\textit{Preprint} \eprint{arXiv:0706.1305
  [gr-qc]})

\bibitem{Aj.al.08}
Ajith P {\em et~al.\/} 2008 {\em Phys. Rev. D\/} {\bf 77} 104017 {E}rratum:
  Phys. Rev. D \textbf{79}, 129901(E) (2009) (\textit{Preprint}
  \eprint{arXiv:0710.2335 [gr-qc]})

\bibitem{Ke.al.09}
Keppel D, Nichols D~A, Chen Y and Thorne K~S 2009  (\textit{Preprint}
  \eprint{arXiv:0902.4077 [gr-qc]})

\bibitem{PrPu.94}
Price R~H and Pullin J 1994 {\em Phys Rev. Lett.\/} {\bf 72} 3297
  (\textit{Preprint} \eprint{arXiv:gr-qc/9402039})

\bibitem{AbPr.96}
Abrahams A~M and Price R~H 1996 {\em Phys Rev. D\/} {\bf 53} 1972
  (\textit{Preprint} \eprint{arXiv:gr-qc/9509020})

\bibitem{Ba.al.02}
Baker J, Campanelli M and Lousto C~O 2002 {\em Phys. Rev. D\/} {\bf 65} 044001
  (\textit{Preprint} \eprint{arXiv:gr-qc/0104063})

\bibitem{So.al.06}
Sopuerta C~F, Yunes N and Laguna P 2006 {\em Phys. Rev. D\/} {\bf 74} 124010
  {E}rrata: Phys. Rev. D \textbf{75}, 069903(E) (2007) \& Phys. Rev. D
  \textbf{78}, 049901(E) (2008) (\textit{Preprint}
  \eprint{arXiv:astro-ph/0608600})

\bibitem{So.al.07}
Sopuerta C~F, Yunes N and Laguna P 2007 {\em Astrophys. J.\/} {\bf 656} L9
  (\textit{Preprint} \eprint{arXiv:astro-ph/0611110})

\bibitem{BuDa.99}
Buonanno A and Damour T 1999 {\em Phys. Rev. D\/} {\bf 59} 084006
  (\textit{Preprint} \eprint{arXiv:gr-qc/9811091})

\bibitem{DaNa.09}
Damour T and Nagar A 2009 The effective one body description of the two-body
  problem {\em Mass and Motion in General Relativity\/} ed Blanchet L,
  Spallicci A and Whiting B (Springer) (\textit{Preprint}
  \eprint{arXiv:0906.1769 [gr-qc]})

\bibitem{Le.al.09}
{Le~Tiec} A, Blanchet L and Will C~M 2010 {\em Class. Quant. Grav.\/} {\bf 27}
  012001 (\textit{Preprint} \eprint{arXiv:0910.4594 [gr-qc]})

\bibitem{Lag}
Lagerstrom P~A 1988 {\em Matched Asymptotic Expansions: Ideas and Techniques\/}
  (New York: Springer)

\bibitem{Bl.al.09}
Blanchet L, Detweiler S, {Le~Tiec} A and Whiting B~F 2010 {\em Phys. Rev. D
  (accepted)\/} (\textit{Preprint} \eprint{arXiv:0910.0207 [gr-qc]})

\bibitem{Bl.al.98}
Blanchet L, Faye G and Ponsot B 1998 {\em Phys Rev. D\/} {\bf 58} 124002
  (\textit{Preprint} \eprint{arXiv:gr-qc/9804079})

\bibitem{Bri.al.09}
Brizuela D, Mart\'in-Garc\'ia J~M and Tiglio M 2009 {\em Phys. Rev. D\/} {\bf
  80} 024021 (\textit{Preprint} \eprint{arXiv:0903.1134 [gr-qc]})

\bibitem{BlIy.03}
Blanchet L and Iyer B~R 2003 {\em Class. Quant. Grav.\/} {\bf 20} 755
  (\textit{Preprint} \eprint{arXiv:gr-qc/0209089})

\bibitem{Sc.al.08}
Schnittman J~D, Buonanno A, van Meter J~R, Baker J~G, Boggs W~D, Centrella J,
  Kelly B~J and McWilliams S~T 2008 {\em Phys. Rev. D\/} {\bf 77} 044031
  (\textit{Preprint} \eprint{arXiv:0707.0301 [gr-qc]})

\bibitem{No.99}
Nollert H~T 1999 {\em Class. Quant. Grav.\/} {\bf 16} R159

\bibitem{MaPo.05}
Martel K and Poisson E 2005 {\em Phys. Rev. D\/} {\bf 71} 104003
  (\textit{Preprint} \eprint{arXiv:gr-qc/0502028})

\bibitem{NaRe.05}
Nagar A and Rezzolla L 2005 {\em Class. Quant. Grav.\/} {\bf 22} R167
  (\textit{Preprint} \eprint{arXiv:gr-qc/0502064})

\bibitem{Gl.al.00}
Gleiser R~J, Nicasio C~O, Price R~H and Pullin J 2000 {\em Phys. Rept.\/} {\bf
  325} 41 (\textit{Preprint} \eprint{arXiv:gr-qc/9807077})

\bibitem{ReWh.57}
Regge T and Wheeler J~A 1957 {\em Phys. Rev.\/} {\bf 108} 1063

\bibitem{Ze1.70}
Zerilli F~J 1970 {\em Phys. Rev. D\/} {\bf 2} 2141

\bibitem{Na.al.03}
Nakano H, Sago N and Sasaki M 2003 {\em Phys. Rev. D\/} {\bf 68} 124003
  (\textit{Preprint} \eprint{arXiv:gr-qc/0308027})

\bibitem{Sa.al.03}
Sago N, Nakano H and Sasaki M 2003 {\em Phys. Rev. D\/} {\bf 67} 104017
  (\textit{Preprint} \eprint{arXiv:gr-qc/0208060})

\bibitem{Ru.al.08}
Ruiz M, Alcubierre M, N{\'u}{\~n}ez D and Takahashi R 2008 {\em Gen. Rel.
  Grav.\/} {\bf 40} 1705 (\textit{Preprint} \eprint{arXiv:0707.4654 [gr-qc]})

\bibitem{Cha}
Chandrasekhar S 1983 {\em The Mathematical Theory of Black Holes\/} (Oxford:
  Oxford University Press)

\bibitem{Som}
Sommerfeld A 1949 {\em Partial Differential Equations in Physics\/} (New York:
  Academic Press)

\bibitem{La1.04}
Lau S~R 2004 {\em J. Comp. Phys.\/} {\bf 199} 376 (\textit{Preprint}
  \eprint{arXiv:gr-qc/0401001})

\bibitem{La2.04}
Lau S~R 2004 {\em Class. Quant. Grav.\/} {\bf 21} 4147 (\textit{Preprint}
  \eprint{arXiv:gr-qc/0401001})

\bibitem{Mi.60}
Misner C~W 1960 {\em Phys. Rev.\/} {\bf 118} 1110

\bibitem{KoSc.99}
Kokkotas K~D and Schmidt B 1999 {\em Living Rev. Rel.\/} {\bf 2} 2
  (\textit{Preprint} \eprint{arXiv:gr-qc/9909058})

\bibitem{Be.al.09}
Berti E, Cardoso V and Starinets A~O 2009 {\em Class. Quant. Grav.\/} {\bf 26}
  163001 (\textit{Preprint} \eprint{arXiv:0905.2975 [gr-qc]})

\bibitem{BrLi.63}
Brill D~R and Lindquist R~W 1963 {\em Phys. Rev.\/} {\bf 131} 471

\bibitem{Gu.al.08}
Gualtieri L, Berti E, Cardoso V and Sperhake U 2008 {\em Phys. Rev. D\/} {\bf
  78} 044024 (\textit{Preprint} \eprint{arXiv:0805.1017 [gr-qc]})

\bibitem{Bl.al.05}
Blanchet L, Qusailah M~S~S and Will C~M 2005 {\em Astrophys. J.\/} {\bf 635}
  508 (\textit{Preprint} \eprint{arXiv:astro-ph/0507692})

\bibitem{Te.72}
Teukolsky S~A 1972 {\em Phys. Rev. Lett.\/} {\bf 29} 1114

\bibitem{Ma.62}
Mathews J 1962 {\em J. Soc. Ind. Appl. Math.\/} {\bf 10} 768

\bibitem{Ze2.70}
Zerilli F~J 1970 {\em J. Math. Phys.\/} {\bf 11} 2203

\bibitem{NePe.66}
Newman E~T and Penrose R 1966 {\em J. Math. Phys.\/} {\bf 7} 863

\bibitem{Go.al.67}
Goldberg J~N, Macfarlane A~J, Newman E~T {\it et al.} 1967
  {\em J. Math. Phys.\/} {\bf 8} 2155

\end{thebibliography}

\end{document}